\documentclass[fleqn,usenatbib]{mnras}

\usepackage{newtxtext,newtxmath}
\usepackage{ulem}
\usepackage[T1]{fontenc}
\usepackage{placeins}
\usepackage{float}
\usepackage{orcidlink}
\usepackage{caption}
\DeclareRobustCommand{\VAN}[3]{#2}
\let\VANthebibliography\thebibliography
\def\thebibliography{\DeclareRobustCommand{\VAN}[3]{##3}\VANthebibliography}

\usepackage{graphicx}	% Including figure files
\usepackage{amsmath}	% Advanced maths commands
\usepackage{multirow} % For multirow cells in tables

\def\mpcoh{\,h^{-1}{\rm Mpc}}

\def\msun{\,{\rm M}_\odot}

\def\m@th{\mathsurround=0pt }
\def\eqalign#1{\null\,\vcenter{\openup1\jot \m@th
 \ialign{\strut\hfil$\displaystyle{##}$&$\displaystyle{{}##}$\hfil
 \crcr#1\crcr}}\,}
 
\title[Galaxy assembly bias in hydro simulations]{Galaxy assembly bias in cosmological hydrodynamical simulations -- a comparison between SIMBA and \textsc{Illustris}TNG 
}

\author[Yang et al.]{
Hong-Gang Yang,$^{\orcidlink{0009-0005-7330-7876}}$$^{1}$\thanks{hong-gang.yang@ed.ac.uk}
 Marcos Pellejero-Ibá$\mathrm{\tilde{n}}$ez,$^{\orcidlink{0000-0003-4680-7275}}$$^{1}$ John A. Peacock,$^{\orcidlink{0000-0002-1168-8299}}$$^{1}$
 Romeel Davé$^{\orcidlink{0000-0003-2842-9434}}$$^{1,2}$\\
$^{1}$Institute for Astronomy, University of Edinburgh, Royal Observatory, Blackford Hill, Edinburgh EH9 3HJ, UK\\
$^{2}$Department of Physics and Astronomy, University of the Western Cape, Bellville, 7535 Cape Town, South Africa}

\date{Accepted XXX. Received YYY; in original form ZZZ}

\pubyear{\the\year{}}

\begin{document}
\label{firstpage}
\pagerange{\pageref{firstpage}--\pageref{lastpage}}
\maketitle

\begin{abstract}
% An accurate description of the galaxy–halo connection is essential for modelling galaxy clustering and interpreting data from ongoing redshift surveys such as DESI. However, understanding the connection beyond halo mass—known as galaxy assembly bias (GAB)—remains a key challenge. 
%\HGY{Galaxy properties in haloes are determined primarily by the halo mass. However, secondary halo properties other than mass may also play a role on e.g., galaxy formation, a phenomenon known as galaxy assembly bias (GAB). Despite extensive studies have explored the correlation of GAB with various secondary halo properties, the physical mechanism causing GAB remains unclear. To shed light on which physical process is responsible for GAB,} 
Modelling of large-scale structure is increasingly concerned with galaxy assembly bias (GAB), the dependence of galaxy clustering on quantities other than halo mass.
We investigate how baryonic physics affects the strength and redshift evolution of GAB using the largest runs of two state-of-the-art cosmological hydrodynamical simulations: SIMBA and \textsc{Illustris}TNG. We quantify GAB by comparing the clustering of stellar-mass-selected galaxies to that of shuffled samples, where galaxies are randomly reassigned to haloes of similar mass. We find that GAB in both simulations increases from approximately zero at $z=5$ to a $\sim$\,$5\%$ change in clustering amplitude at $z=2$. After this epoch, the trends diverge: GAB in TNG continues to increase, reaching $\sim$\,$10\%$ at $z=0$, while in SIMBA it decreases to nearly zero. By further shuffling galaxies within bins of halo mass and cosmic environment -- characterised by smoothed matter overdensity ($\delta_5$) and tidal anisotropy ($\alpha_5$) -- we show that most of the GAB in both simulations can be attributed to the overdensity, while tidal anisotropy contributes negligibly in both simulations. 
Exploring this effect from the point of view of the halo occupation distribution (HOD), we find that numbers of central and satellite galaxies vary with overdensity -- but only near the respective turn-on masses for these two constituents: the galaxy contents of high-mass haloes are very nearly independent of environment. We present a simple parameterisation that allows the HOD modelling to be modified to reflect this form of density-dependent GAB.
%; this should be useful in the construction of mock galaxy catalogues. 
%\HGY{Our results show that GAB is closely linked to baryonic processes in hydrodynamical simulations, representing a step towards a better understanding of its physical drivers and, consequently, towards more accurate modelling.}

%Consistent with this, halo occupation distribution (HOD) measurements reveal that GAB arises when low-mass haloes in overdense regions are more likely to host galaxies than those in underdense regions. Our findings demonstrate that GAB is sensitive to the baryonic physics implemented in simulations and provide insight into its physical origin, paving the way for more accurate models of galaxy clustering. \JAP{Last sentence needs to be more specific. Can we perhaps say something about using this as a prior for the range of possible GAB effects in mock catalogues?}
\end{abstract}


\section{Introduction}
The spatial distribution of galaxies is a key observable for cosmological studies, as it traces the underlying large-scale structure (LSS) of the Universe, which is primarily composed of invisible dark matter. In the framework of hierarchical structure formation, galaxies form within dark matter haloes that originate from primordial density fluctuations and grow through gravitational collapse, accretion, and mergers \citep{White_1978,Blumenthal_1984,Lacey_Cole_1993}. A precise understanding of the galaxy–halo connection is therefore essential for interpreting data from current and forthcoming galaxy redshift surveys, such as the Dark Energy Spectroscopic Instrument (DESI; e.g., \citealp{DESI_overview,DESI_DR2_BAO,DESI_DR1_BAO,DESI_DR1_fs}) and the Euclid mission (e.g., \citealp{Euclid_EWS,Euclid_Q1_overview}).

One of the earliest approaches to modelling the galaxy–halo connection is the halo occupation distribution (HOD; e.g., \citealp{Benson_2000, Seljak_2000, Peacock_2000, White_2001, Berlind_2002, COORAY_2002}), which describes the average number of galaxies hosted by a dark matter halo as a function of its mass. The simplest form of the HOD assumes that halo mass alone determines galaxy occupancy, a notion motivated by excursion set theory \citep{press_schechter,excursion_set}, which predicts that halo properties beyond mass are uncorrelated with the cosmic environment due to its random-walk formalism \citep{Lemson_Kauffmann_1999}. However, this assumption was later challenged by \citet{Sheth_2004}, who demonstrated using $N$-body simulations that haloes in dense environments tend to form earlier than haloes of the same mass in less dense regions. Building on this, \citet{Gao_2005} used the Millennium Simulation \citep{Millennium} to show that halo clustering depends not only on mass but also on halo formation time. This secondary dependence was subsequently extended to other halo properties, such as concentration and spin \citep{Wechsler_2006, Gao_2007, Angulo_2008,Salcedo_2018,Sato_Polito_2019} and the cosmic web environment \citep{Borzyszkowski_2017,Tojeiro_2017,Yang_2017,Paranjape2018,Han_2018,Ramakrishnan_2019}. A dependence of halo clustering on properties beyond their mass is commonly referred to as halo assembly bias (HAB; e.g. \citealp{Wechsler_2018}), although the secondary properties may not correspond exactly to the halo assembly history.

If galaxy formation is influenced by the secondary halo properties that give rise to HAB, then the core assumption of the simplest HOD model -- that galaxy occupancy depends solely on halo mass -- is violated. This violation can systematically alter predictions of galaxy clustering. The phenomenon in which the galaxy clustering depends on halo properties beyond mass is known as galaxy assembly bias (GAB, see \citealp{Wechsler_2018} for a review).

The existence of GAB in simulation-based catalogues was confirmed shortly after the discovery of HAB. \citet{Zhu_2006} showed that galaxy properties and occupancy correlate with secondary halo properties. \citet{Croton_2007} introduced the shuffling technique for quantifying GAB, and used it to demonstrate that galaxy clustering is sensitive to halo properties beyond mass. Since then, numerous studies have examined the impact of GAB on galaxy clustering and the degree to which it is driven by specific halo internal properties such as age, concentration, or spin \citep{Zu_2008,Chaves_2016,Zehavi_2018,Artale_2018,Contreras_2019,Xu_2021,Moreno_2025} -- or by environmental properties such as smoothed overdensity and tidal anisotropy \citep{Zehavi_2018,Artale_2018,Xu_2021,Alam_2023,Montero_2024,Wang_2024,Perez_2024,Lacerna2025}. These investigations have primarily relied on galaxy catalogues generated from cosmological hydrodynamical simulations (mainly \textsc{Illustris}TNG; see e.g. \citealp{MD2020,MD2021,Xu_2020}) and/or semi-analytical galaxy formation models applied to $N$-body simulations (e.g., \citealp{Contreras_2019}).

Despite this extensive body of work, the physical nature of GAB remains unclear and its manifestation in observational data is still debated: different studies disagree over whether the driving factor is halo concentration, or on whether any GAB effect applies both to central and satellite galaxies (e.g., \citealp{Vakili_2019,Yuan_2021,Pearl_2024,Zhang_2025}). But if GAB is indeed present in real data, then halo mass alone must be insufficient to fully describe the galaxy–halo connection in the era of precision cosmology, biasing the predicted isotropy or scale dependence of galaxy clustering (e.g. \citealp{Wu_2008,McCarthy_2019,Obuljen_2019,Hadzhiyska_2020,Sergio_2023}). Understanding the underlying physical mechanism of GAB is therefore both critical and timely for interpreting data from galaxy redshift surveys.

In this paper, we propose that comparing the strength and evolution of GAB across different cosmological hydrodynamical simulations can provide valuable insight into its physical origin. Previous studies have shown that GAB signals in \textsc{Illustris}TNG and in semi-analytical models are consistent (e.g., \citealp{Hadzhiyska_2021}), but direct comparisons between different hydrodynamical simulations have not been made. Although modern hydrodynamical simulations broadly reproduce observed galaxy properties, they adopt different subgrid prescriptions for baryonic physics -- particularly stellar and AGN feedback -- which can lead to substantial differences in the displacement of gas particles, halo baryon fraction, the baryonic imprint on the matter power spectrum, and star formation history (see e.g., studies of the CAMELS project; \citealp{CAMELS_VN,CAMELS_Ni,CAMELS_Iyer}).
% differ significantly in aspects such as gas heating efficiency and small-scale clustering (e.g., \citealp{CAMELS_Ni}). 
Comparing GAB across simulations with differing baryonic prescriptions can thus help isolate the physical processes responsible for the emergence of GAB.

This paper is structured as follows. In Section~\ref{sec_data}, we introduce the two hydrodynamical simulations used and describe the galaxy selection criteria. Section~\ref{Sec_GAB} outlines our methodology for measuring GAB and presents its redshift evolution, along with an assessment of the contributions from smoothed overdensity and tidal anisotropy. In Section~\ref{Sec_HOD}, we examine GAB through the lens of the HOD. Section~\ref{Sec_lmh} quantifies the contributions of low-mass haloes to the total GAB. In Section~\ref{Sec_fit}, we provide a fitting formulation to capture how does HOD vary from the average according to the cosmic environment. We summarise and discuss our findings in Section~\ref{conclusion}.

\section{Data}\label{sec_data}
We use two state-of-the-art cosmological hydrodynamical simulations to study the redshift evolution of GAB: SIMBA \citep{SIMBA} and \textsc{Illustris}TNG \citep{TNG_DR}. These simulations differ in their $N$-body and hydrodynamics solvers, star-formation criteria, super-massive black hole (SMBH) seeding and growth prescriptions, and most notably in their implementations of stellar and AGN feedback. These differences can lead to substantially different halo baryon fractions and intergalactic medium thermal states, which in turn can affect galaxy formation and evolution. 
% These simulations include physical prescriptions for various baryonic processes, such as stellar winds, gas cooling, and AGN feedback. Importantly, the baryonic physics implemented in SIMBA and TNG differ in key aspects, such as the treatment of gas heating by feedback (see, e.g., \citealp{CAMELS_Ni}). 
Despite these differences, both simulations reproduce the observed galaxy stellar mass function at $z=0$ with high fidelity. Comparing the GAB evolution in the two simulations thus provides an opportunity to gain insights into the physical mechanisms driving GAB. Below, we briefly summarise the relevant features of each simulation.

\subsection{SIMBA}
The SIMBA simulation suite models galaxy formation within a cosmological context using a modified version of the GIZMO code \citep{GIZMO,Hopkins_2017} with the meshless finite mass (MFM) hydrodynamics solver. Stellar feedback is modelled as galactic winds. The speed of the wind particles scales with galaxy circular velocity, while the temperature of the wind particles is set by the supernovae energy minus wind kinetic energy. AGN feedback from black holes has two modes: high mass loading outflows in the radiative `QSO' mode, and lower mass loading but faster outflows in the bipolar jet mode at low Eddington ratio (see \citealp{SIMBA, CAMELS_VN} and references therein).

% SIMBA incorporates a range of baryonic feedback processes, including stellar feedback, AGN feedback in the form of radiative winds and collimated jets, and X-ray feedback from black hole accretion \citep{SIMBA}.

In this work, we use the largest SIMBA run ({\tt m100n1024}), which evolves $1024^3$ dark matter and $1024^3$ baryonic particles in a cubic box of side $100\mpcoh$. The mass resolution is $9.6\times10^7\,\mathrm{M}_{\odot}$ for dark matter particles and $1.82\times10^7\,\mathrm{M}_{\odot}$ for baryonic particles. The simulation assumes the \textit{Planck} 2015 cosmological parameters \citep{Planck2015} with specifications as follows: $\Omega_{\Lambda,0} = 0.7$, $\Omega_{\mathrm{m},0} = 0.3$, $\Omega_{\mathrm{b},0} = 0.048$, $\sigma_8 = 0.82$, $n_s = 0.97$, and $H_0 = 68\,\mathrm{km\,s^{-1}\,Mpc^{-1}}$.

Haloes are identified on the fly using a 3-D friends-of-friends (FOF) algorithm with linking length $b=0.2$ times the typical interparticle spacing, and galaxies are identified  from star particles using FOF with a spatial linking length of 0.0056 times the typical interparticle spacing \citep{SIMBA}. 
%we adopt the total FOF group mass as the halo mass definition. 
The minimum stellar mass for resolved galaxies in SIMBA is $5.8\times10^{8}\,\mathrm{M}_\odot$. All galaxies with stellar mass greater than $5.8\times10^{8}\,\mathrm{M}_\odot$ in the SIMBA catalogue are used in our analysis to maximise statistical power. The number density of galaxies corresponding to this stellar mass threshold varies with redshift, being $0.043\,h^3\,\mathrm{Mpc}^{-3}$ at $z=0$ and $0.019\,h^3\,\mathrm{Mpc}^{-3}$ at $z=2$.
%\sout{To ensure that the resolution is reliable for our study, we checked that changing the stellar mass threshold for galaxies in SIMBA to $1.2\times10^{9}\,\mathrm{M}_\odot$ do not alter our conclusions qualitatively.}

\subsection{\textsc{Illustris}TNG}

The \textsc{Illustris}TNG project comprises a suite of large-volume cosmological magnetohydrodynamical simulations \citep{TNG_Method_1,TNG_Method_2} based on the moving-mesh code AREPO \citep{AREPO}. Galactic winds driven by stellar feedback are modelled kinetically as decoupled particles that are stochastically and isotropically ejected from star-forming gas, with only 10\% of the supernovae energy turns into thermal energy. AGN feedback has three modes: thermal, kinetic, and radiative. Thermal mode at high accretion rate heats up the feedback sphere; kinetic mode at low accretion rate injects kinetic energy in a random direction into the feedback sphere; and the radiative mode modifies the heating and cooling of the gas surrounding the SMBH (see \citealp{TNG_Method_1,TNG_Method_2, CAMELS_VN} and references therein).

We use the largest run of the \textsc{Illustris}TNG project, namely TNG300-1 \citep{TNG_DR,Pillepich_2017}, which we refer to as TNG hereafter for simplicity. TNG simulates a cubic volume with $L_{\mathrm{box}} = 205\mpcoh$ using $2500^3$ particles for both dark matter and baryons. The mass resolution for baryon and dark matter are $7.6\times10^6\,h^{-1}\,\mathrm{M_\odot}$ and $4.0\times10^7\,h^{-1}\,\mathrm{M_\odot}$, respectively. TNG also adopts the \textit{Planck} 2015 cosmological parameters \citep{Planck2015}: $\Omega_{\Lambda,0} = 0.6911$, $\Omega_{\mathrm{m},0} = 0.3089$, $\Omega_{\mathrm{b},0} = 0.0486$, $\sigma_8 = 0.8159$, $n_s = 0.9667$, and $h =0.6774 $.

Haloes in TNG are identified using the FOF algorithm with linking length $b=0.2$. Galaxies are defined here as subhaloes in TNG with stellar mass greater than $10^8\,\mathrm{M_\odot}$; these are identified within gravitationally bound substructures using the \texttt{Subfind} algorithm \citep{Subfind}. 
%We adopt the total FOF mass as the halo mass definition. 
To ensure a clean sample, galaxies with \texttt{SubhaloFlag} = 0 (typically considered to be non-cosmological or spurious) are excluded.

To ensure a fair comparison with SIMBA, we select galaxies in TNG by stellar mass to match the number density of galaxies in SIMBA. At $z=0$, this corresponds to a stellar mass threshold of $4.5\times10^8\,\mathrm{M}_{\odot}$. The exact value varies slightly (between $2.9\times10^8\,\mathrm{M}_{\odot}$ and $1.2\times10^9\,\mathrm{M}_{\odot}$) with redshift.

\begin{figure*}
	\includegraphics[width=17.5cm]{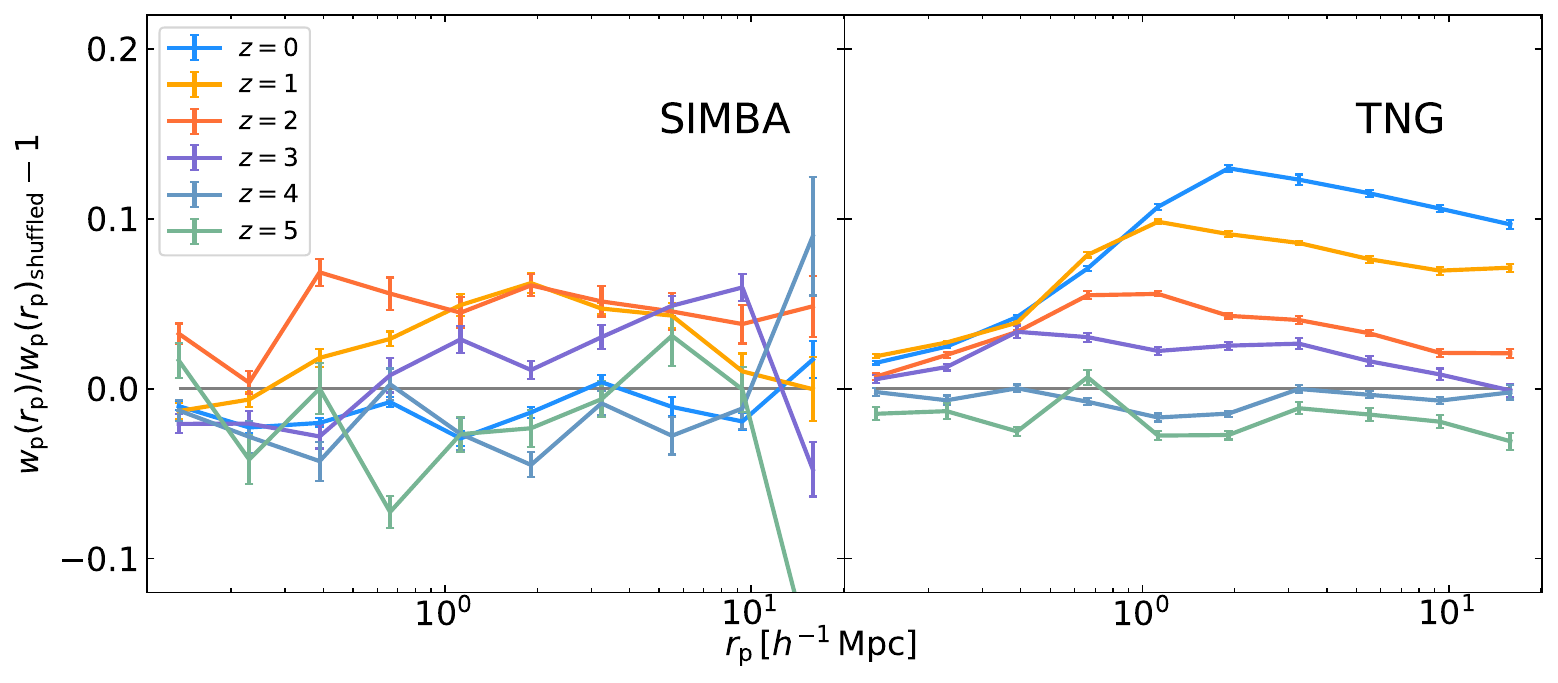}
    \caption{The effect of galaxy assembly bias on galaxy clustering, as quantified by  $[w_{\mathrm{p}}(r_{\mathrm{p}})/w_{\mathrm{p}}(r_{\mathrm{p}})_{\mathrm{shuffled}}-1]$ at redshifts $z=5$, $4$, $3$, $2$, $1$, and $0$. The left panel shows the results for SIMBA; the right panel for TNG. Error bars indicate $1\sigma$ uncertainties computed using Eq.~(\ref{error}).}
    \label{bias_evolution}
\end{figure*}

\section{The evolution of galaxy assembly bias}\label{Sec_GAB}

%However, the physical origin of GAB remains unclear. A deeper understanding of its physical drivers and redshift evolution is crucial for advancing theories of galaxy formation and for accurately interpreting data from current and forthcoming galaxy redshift surveys. One promising approach is to study 
% Here, we focus on the redshift evolution of GAB, which may offer valuable insights into its physical drivers. 

Past studies of galaxy assembly bias (GAB) have primarily concentrated on quantifying its significance and its correlation with secondary halo properties at a single epoch. Studying the evolution of GAB can offer valuable insights. For example, 
\citet{Contreras_2019} investigated this evolution using a semi-analytical galaxy formation model applied to the Millennium-WMAP7 N-body simulation \citep{Guo_2013}, finding that GAB increases monotonically with redshift. \citet{Sergio_2021} measured the evolution of GAB for galaxy samples from TNG, a semi-analytical model, and the subhalo abundance matching technique, finding differences in both magnitude and evolution across these samples. 

Here, we investigate the redshift evolution of GAB in two hydrodynamical simulations, SIMBA and TNG. Hydrodynamical simulations incorporate a wide range of astrophysical processes that might affect galaxy formation, and therefore may affect both the strength and evolution of GAB. Notably, SIMBA and TNG implement very different physical models for feedback processes, whose effects have been systematically studied by the CAMELS project (e.g., \citealp{CAMELS_VN,CAMELS_Ni}). By studying how differences in baryonic physics lead to similarities or discrepancies in GAB evolution, we thus have the potential to identify which physical processes give rise to GAB. As a first step, our focus here is on the simpler initial question of the empirical form of GAB found in these simulations.

In this section, we measure the strength and the evolution of the total GAB from both SIMBA and TNG, looking in particular at its correlation with cosmic environment as represented by matter overdensity and tidal anisotropy.

\subsection{The total GAB}\label{all_GAB}

To quantify the strength of GAB, we apply the shuffling technique introduced by \citet{Croton_2007}, in which GAB is measured as the ratio of galaxy clustering statistics before and after shuffling. Specifically, haloes are divided into subsamples based on halo mass -- defined as the total mass of all particles in the halo -- using logarithmic bins of width $0.1\,\mathrm{dex}$. We have verified that modest changes in bin width do not significantly affect our results. During the shuffling procedure, galaxies from one halo are randomly reassigned to another halo (which might or might not have originally hosted galaxies) within the same mass bin. The central galaxy retains its relative position with respect to the halo coordinate (both galaxy and halo coordinates are defined by the position of the most-bound particle), while satellite galaxies retain their relative positions to the central galaxy.  This procedure removes any dependence of the halo-galaxy connection on halo properties other than mass, including both the halo internal properties (e.g., formation time, concentration) and environmental properties (e.g., local overdensity, tidal anisotropy). The shuffled galaxy catalogue therefore contains no GAB, and comparing clustering measurements before and after shuffling isolates the effect of GAB on galaxy clustering. 

By construction, this shuffling process preserves the mass-dependent trends that affect galaxy clustering: the overall HOD and the radial distribution of galaxies within haloes.
However, the galaxy distribution within haloes may be anisotropic: aligning with the halo shape, which is itself influenced by its cosmic web environment. The shuffling does not preserve such alignments. It is therefore possible that any detected GAB signal might arise purely from the location-dependent orientations of haloes, rather than via more complex astrophysics. We have tested this possibility in a simple way by randomly rotating the galaxy contents of each halo; the resulting change in clustering is typically $\lesssim 1\%$, which is negligible compared to the GAB signal and can thus be safely ignored.

\begin{figure*}
	\includegraphics[width=1.4\columnwidth]{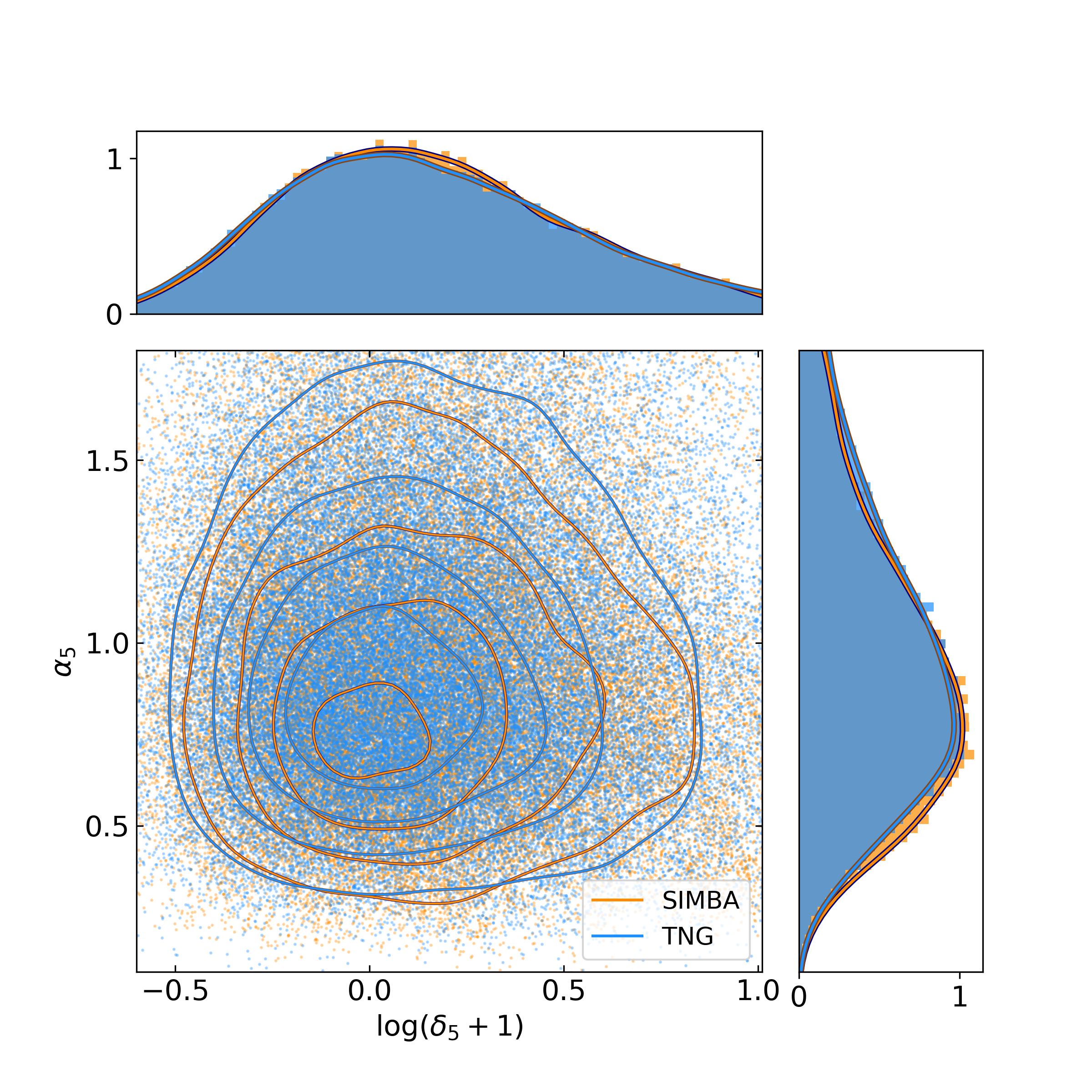}
    
    \caption{The distribution of haloes on the $\delta_5-\alpha_5$ plane. Each orange (blue) dots is one halo from the SIMBA (TNG) sample at $z=0$. The contours indicate the 2-dimensional probability distribution of galaxies on the $\delta_5-\alpha_5$ plane. The marginalized distributions for $\delta_5$ and $\alpha_5$ are shown in the top and right panels, respectively.}
    \label{fig1}
\end{figure*}
We choose the projected correlation function $w_{\mathrm{p}}(r_{\mathrm{p}})$ as the galaxy clustering statistic; 
%\sout{(we checked that changing the galaxy clustering statistic to two-point autocorrelation function does not qualitatively alter our conclusions)}, 
this is defined via
\begin{equation}
    w_{\mathrm{p}}(r_{\mathrm{p}})=2 \int_0^{\pi_{\mathrm{max}}}\xi(r_{\mathrm{p}},\pi)\,\mathrm{d}\pi,
\end{equation}
where $\xi(r_{\mathrm{p}},\pi)$ is the two-dimensional correlation function along the projected and line-of-sight directions $r_{\mathrm{p}}$ and $\pi$, respectively. We calculate $w_{\mathrm{p}}(r_{\mathrm{p}})$ in 10 logarithmically spaced bins between $0.1\mpcoh$ and $20\mpcoh$ with $\pi_{\mathrm{max}}=30\mpcoh$. The computation of $w_{\mathrm{p}}(r_{\mathrm{p}})$ made use of the \texttt{halotools} package \citep{halotools}. To examine the evolution of GAB, we compute $w_{\mathrm{p}}(r_{\mathrm{p}})$ before and after shuffling at $z=5,4,3,2,1,0$. We quantify the effect of GAB using $[w_{\mathrm{p}}(r_{\mathrm{p}})/w_{\mathrm{p}}(r_{\mathrm{p}})_{\mathrm{shuffled}}-1]$. To reduce the noise introduced by the shuffling procedure, we repeat the shuffling ten times and define the GAB signal as the average of the quantity $[w_{\mathrm{p}}(r_{\mathrm{p}})/w_{\mathrm{p}}(r_{\mathrm{p}})_{\mathrm{shuffled}} - 1]$ across the ten realisations. A positive value indicates that shuffling reduces galaxy clustering, implying that the clustering amplitude in the original data is enhanced by GAB.

To estimate the uncertainty in our results, we divide both the original and shuffled simulation boxes into $5\times5$ equal-sized subboxes in the $X-Y$ plane. We compute the quantity $[w_{\mathrm{p}}(r_{\mathrm{p}})/w_{\mathrm{p}}(r_{\mathrm{p}})_{\mathrm{shuffled}}-1]$ while excluding one subbox at a time. The resulting jackknife variance quantifies the uncertainty due to the finite volume of the simulation and is denoted as $\sigma^2_{\mathrm{cosmic}}$. An additional source of uncertainty arises from the stochastic nature of the shuffling procedure. This is quantified by repeating the shuffling process ten times and computing the variance in GAB across the realisations, denoted as $\smash{\sigma^2_{\mathrm{shuffle}}}$. For each shuffle realisation, we compute $\sigma^2_{\mathrm{cosmic},i}$ and the final value of $\smash{\sigma^2_{\mathrm{cosmic}}}$ is taken to be the average over the ten realisations. The total uncertainty is then given by
\begin{equation}
    \sigma = \sqrt{\sigma^2_{\mathrm{cosmic}}+\sigma^2_{\mathrm{shuffle}}}\label{error}.
\end{equation}

The evolution of the total GAB in SIMBA and TNG is shown in Fig.~\ref{bias_evolution}. In SIMBA, the GAB is approximately $-3\%$ around $1\,h^{-1}\mathrm{Mpc}$ at $z=5$, increases with decreasing redshift, and reaches a peak of $\sim5$\,$\%$ at $z=2$. After $z=2$, the GAB declines, yielding a value of almost zero at $z=0$. The TNG simulation shows a GAB close to zero at $z=5$, which increases steadily with time, reaching $\sim$\,$5\%$ at $z=2$, in agreement with SIMBA at that epoch. The GAB in TNG continues to rise beyond $z=2$, attaining a maximum of $\sim$\,$10\%$ at $z=0$. This GAB evolution in TNG is similar to the results of \citet{Contreras_2019} when galaxies are selected by stellar mass, using a semi-analytical galaxy formation model. 

The consistent GAB trend in both simulations prior to the cosmic noon epoch ($z \sim 2$), followed by divergent behaviour at lower redshifts, is one of our key findings. In both SIMBA and TNG, this redshift corresponds to the onset of substantial AGN feedback that causes galaxies to quench; prior to this redshift, feedback is dominated by stellar energy input.  This suggests that AGN feedback is a key difference between these models that drives differences in the GAB and thus galaxy clustering.  Indeed, AGN feedback in different models has been shown to have dramatically different impacts on the circum-galactic medium in massive haloes.  We have seen that AGN feedback in SIMBA is included as stably bipolar and decoupled jets, whereas in IllustrisTNG the jet direction is randomized at every timestep and not decoupled.  The result is that SIMBA produces significantly lower baryon fractions particularly in group-scale haloes~\citep{Appleby_2021,Oppenheimer_2021,Sorini_2022}, and transports gas up to $\sim10\mpcoh$ away from its origin \citep{Borrow_2020,CAMELS_Ni}.  This has an impact within the CGM on for instance the Sunyaev-Zeldovich $y$ decrement \citep{Yang_2022} and the intergalactic medium as traced by the Lyman alpha forest \citep{Tillman_2023}.  Our work here suggests that there may also be a significant impact on GAB and hence on galaxy clustering.  We leave a more detailed investigation into the exact process by which AGN feedback impacts the GAB for future work.

% \HGY{Our results thus show that differences in baryonic implementations across hydrodynamical simulations can significantly affect the GAB and thus the galaxy clustering.}  Furthermore, this divergence coincides with the peak of the cosmic star formation rate, when feedback processes such as stellar and AGN feedback are most influential. \HGY{Compared to TNG, the AGN feedback implemented in SIMBA is substantially more efficient at transporting energy out to the circumgalactic medium \citep{Yang2024} and the intergalactic medium \citep{Tillman_2023}. This can drive gas displacements of up to $\sim10\mpcoh$ \citep{CAMELS_Ni}, thereby influencing the baryonic content of neighbouring haloes.} 
% We suspect that the difference in behaviour is tied in particular to the more explicit treatment of jets in SIMBA, which have a characteristic impact on the circumgalactic medium \citep{Yang2024}. 
% \HGY{We suspect that the difference in the GAB evolution is tied in particular to the more explicit treatment of jets in SIMBA, but the detailed exploration of this issue is deferred to future work. }

\subsection{GAB from the cosmic environment}\label{env_GAB}
\begin{figure*}
    \centering

    \includegraphics[width=17.5cm]{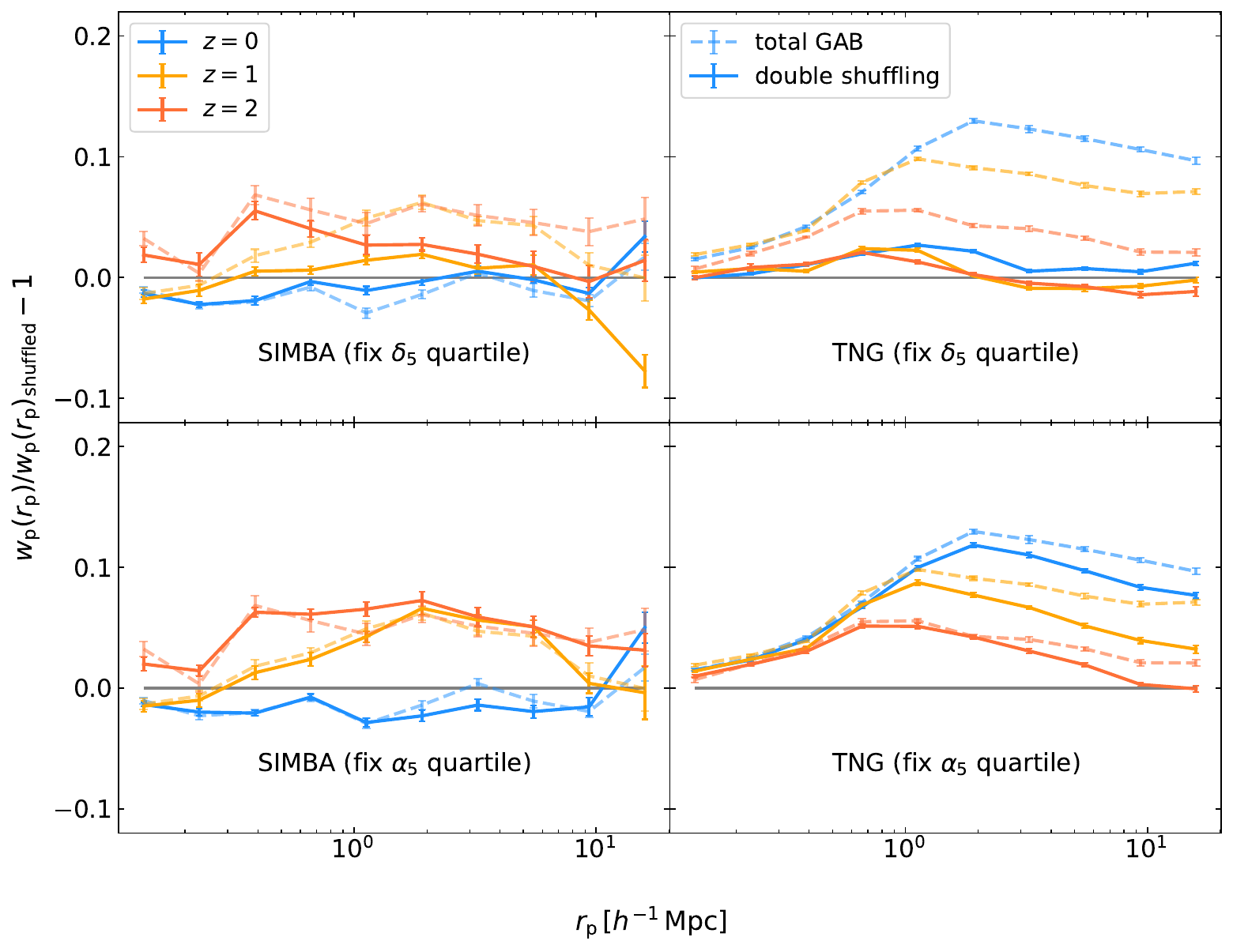}

    \caption{Comparison between the total GAB (dashed lines; shuffled by halo mass only) and the GAB with cosmic environment dependence removed (solid lines; shuffled by both halo mass and [$\delta_5$ or $\alpha_5$] ) at $z=2,1,0$. Left and right panels show results from SIMBA and TNG, respectively. Top and bottom panels correspond to cosmic environment defined by $\delta_5$ and $\alpha_5$, respectively. The difference between dashed and solid curves reflects the contribution of the corresponding cosmic environment variable to the GAB.}
    \label{double_bin}
\end{figure*}
The above exercise in halo shuffling has demonstrated that GAB exists in these hydrodynamical simulations: there must thus be some property of the environment inhabited by a given halo that influences its galaxy contents. The environment has traditionally been quantified in terms of local properties (the density contrast) and non-local ones (the tidal field), and we now explore the extent to which these properties influence the GAB signal that we have detected.

% \HGY{In the previous subsection, we showed that the GAB evolution trends differ across hydrodynamical simulations, suggesting a potential connection between GAB and baryonic physics, such as the strength of feedback. Feedback strength itself varies with cosmic environment: haloes located near cosmic knots are more strongly affected by feedback processes than those near cosmic voids. Motivated by this, we now investigate how the cosmic environment of haloes contributes to the total galaxy assembly bias. In particular, we focus on two environmental properties: the matter overdensity $\delta$ and the tidal anisotropy $\alpha$.}

To compute the overdensity field $\delta$, we assign all massive particles in the simulation box to a $256^3$ grid using the nearest grid point (NGP) interpolation scheme. The resulting density field is then smoothed with a real-space top-hat filter, to capture the large-scale environment surrounding each halo rather than the density internal to it. The filtering radius is of course arbitrary, but we adopt the common choice of $5\mpcoh$, which is substantially beyond the virial radius of any halo, while still retaining information on the cosmic web. We checked that changing the smoothing radius to $3,7,9\,\mpcoh$ makes no qualitative difference to the behaviour. We denote the smoothed overdensity as $\delta_5$. Each halo is then assigned the value of $\delta_5$ from the grid cell in which its centre resides, representing the local cosmic overdensity of its environment.

To quantify the tidal environment, we first define the dimensionless potential field $\widetilde{\Phi}$ through the Poisson equation
\begin{equation}
    \nabla^2 \widetilde{\Phi} = \delta,
\end{equation}
from which we can define the tidal tensor
\begin{equation}
    T_{ij}=\frac{\partial^2\tilde{\Phi}}{\partial x_i\partial x_j}.
\end{equation}
We then follow \cite{Paranjape2018} and define the tidal anisotropy as follows:
\begin{equation}
    \alpha_R = \sqrt{q_R^2}/(1+\delta_R)^\beta,
    \label{ani_def}
\end{equation}
where
\begin{equation}
    q^2 = (\lambda_1-\lambda_2)^2+(\lambda_2-\lambda_3)^2+(\lambda_3-\lambda_1)^2,
\end{equation}
with $\lambda_1,\lambda_2,\lambda_3$ being the eigenvalues of the tidal tensor $T_{ij}$, $R$ is the smoothing scale set to be $5\mpcoh$, and $\beta$ is defined so as to minimise the correlation between $\delta_R$ and $\alpha_R$, so that any potential galaxy-halo connection dependence on $\alpha_R$ does not stem from $\delta_R$. To determine the value of $\beta$, we measure $\alpha_R$ and $\delta_R$ for all the haloes with $M_\mathrm{h}>10^{10.4}\,\mathrm{M}_\odot$ and derive the value of $\beta$ that minimises their Spearman rank correlation coefficient. To illustrate the lack of correlation, we show the joint distribution of $\alpha_R$ and $\delta_R$ for haloes in SIMBA and TNG at $z=0$ in Fig.~\ref{fig1}. The circular shapes of the contours demonstrate that $\delta_5$ and $\alpha_5$ of haloes are not correlated with each other. Notice that $\beta$ needs to be calculated separately for different simulations and at different redshifts.

To investigate how much of the GAB can be attributed to the cosmic environment as described by $\delta_5$ or $\alpha_5$, we employ a double-shuffling technique, during which galaxies are reassigned to haloes within the same halo mass bin as well as the $\delta_5$ or $\alpha_5$ bin. To be specific, we rank haloes in the same halo mass bin by their $\delta_5$ or $\alpha_5$ and divide the subsample into 4 quartiles accordingly. The reassigning of the galaxy contents of haloes are then performed within the same $\delta_5$ or $\alpha_5$ quartile for each halo mass bin. We now define the double-property (i.e. halo mass and [$\delta_5$ or $\alpha_5$]) shuffled sample as the shuffled catalogue. This catalogue contains no GAB other than that correlated with $\delta_5$ or $\alpha_5$. The additional GAB signal in the original sample, quantified by $[w_{\mathrm{p}}(r_{\mathrm{p}})/w_{\mathrm{p}}(r_{\mathrm{p}})_{\mathrm{shuffled}} - 1]$, would therefore reduce to zero if the double-property shuffled sample retained the same level of GAB as the original sample. Conversely, if no GAB were associated with $\delta_5$ or $\alpha_5$, the double-property shuffled sample would contain no GAB, and the resulting $[w_{\mathrm{p}}(r_{\mathrm{p}})/w_{\mathrm{p}}(r_{\mathrm{p}})_{\mathrm{shuffled}} - 1]$ would be identical to that obtained from the halo-mass-only shuffled sample.

% \begin{figure*}
%     \centering
%     \includegraphics[width=1\linewidth]{figs/HOD_combined.pdf}
%     \caption{HOD curves measured from the SIMBA and TNG simulations at $z=0$ (\textit{left}) and $z=2$ (\textit{right}). The total HODs of all galaxies for SIMBA and TNG are shown as black and grey solid lines, respectively. Each total HOD is decomposed into contributions from central galaxies (orange for SIMBA, blue for TNG) and satellite galaxies (red for SIMBA, purple for TNG). Solid lines represent measurements from all haloes in the sample, while dashed (dotted) lines correspond to haloes in the top (bottom) 25\% of the cosmic environment distribution, quantified by $\delta_5$ (\textit{upper}) or $\alpha_5$ (\textit{lower}), in the corresponding halo mass bin. In each panel we present the ratio of each subsample’s HOD to the corresponding average HOD at the bottom, highlighting deviations from the mean.}
%     \label{HOD_0_delta}
% \end{figure*}
\begin{figure*}
    \centering
    \includegraphics[width=1\linewidth]{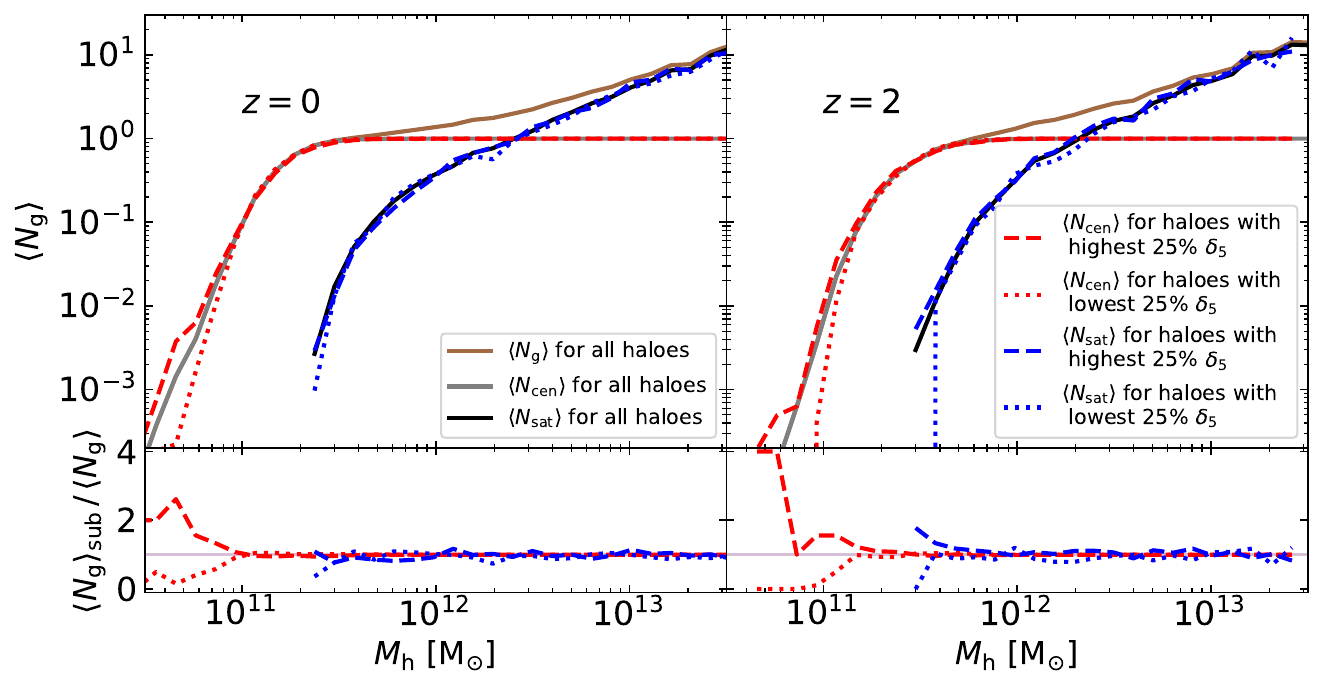}
    \caption{HOD curves measured from SIMBA at $z=0$ (\textit{left}) and $z=2$ (\textit{right}). The total HOD of all galaxies for SIMBA is shown as the brown solid line. The total HOD is decomposed into contributions from central galaxies (gray line) and satellite galaxies (black line). Solid lines represent measurements from all haloes in the sample, while dashed (dotted) lines correspond to haloes in the highest (lowest) 25\% of the cosmic environment distribution, quantified by $\delta_5$ in the corresponding halo mass bin. In each panel we present the ratio of each subsample’s HOD to the corresponding total HOD at the bottom, highlighting deviations from the mean.}
    \label{simba_delta}
\end{figure*}

\begin{figure*}
    \centering
    \includegraphics[width=1\linewidth]{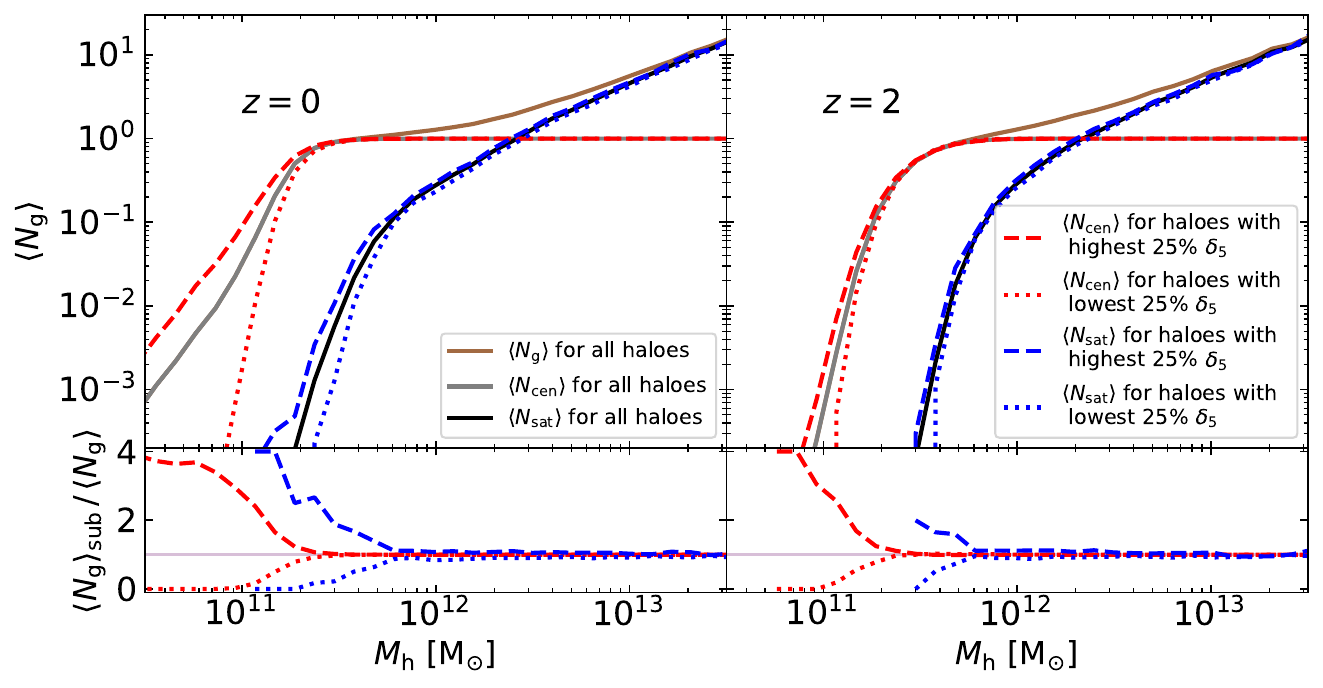}
    \caption{The same as Fig.~\ref{simba_delta}, but for TNG.}
    \label{tng_delta}
\end{figure*}
The results for the double shuffling are shown in Fig. \ref{double_bin}. Since the total GAB at $z=5,4,3$ is negligible in both SIMBA and TNG, we omit these redshifts from the plot. 
%\sout{We first focus on the top panels, which correspond to shuffling by halo mass and $\delta_5$. In SIMBA, we find that at higher redshifts ($z=2,3$), $\delta_5$ accounts for approximately half of the total GAB. At lower redshifts ($z=0,1$), although the GAB is reduced to zero after double shuffling, the total GAB from halo-mass-only shuffling is already not evident. We therefore conclude that in SIMBA, roughly half of the GAB originates from the $\delta_5$ dependence of the halo–galaxy connection.} 
From the top panels of Fig.~\ref{double_bin}, we see that in both SIMBA and TNG, $\delta_5$ accounts for most of the GAB across all redshifts.
The results for double shuffling with $\alpha_5$ are shown in the bottom panels of Fig.~\ref{double_bin}. In SIMBA, the contribution of $\alpha_5$ to the total GAB is negligible at all redshifts. 
Thus, we conclude that $\alpha_5$ does not contribute to GAB in SIMBA. In contrast, for TNG, $\alpha_5$ contributes a small but non-negligible fraction of the total GAB at all redshifts. However, this contribution remains minor compared to that from $\delta_5$. Notice that a dependence of GAB on $\delta_5$ implies that halo assembly bias (HAB) must itself depend on $\delta_5$, as demonstrated in previous studies (e.g. \citealp{Dalal_2008,Tucci_2020,Mansfield_2020}). When this HAB dependence is combined with a tendency for galaxy formation to vary with $\delta_5$, the result is a GAB dependence on $\delta_5$. We would further explore this tendency in the next Section.

To conclude this section, we summarise our main findings on the redshift evolution of GAB, as measured from the hydrodynamical simulations SIMBA and TNG:
\begin{enumerate}
\item In both SIMBA and TNG, GAB is negligible at $z=5$ and gradually increases, reaching a strength of approximately 5\% by $z=2$.
\item After $z=2$, the GAB evolution diverges: in SIMBA, the signal declines to nearly zero by $z=0$, whereas in TNG, it continues to increase, reaching about 10\% at $z=0$.
\item The cosmic environment, as quantified by the smoothed overdensity $\delta_5$, accounts for the majority of the total GAB signal. In contrast, the contribution from tidal anisotropy $\alpha_5$ is relatively minor.
\end{enumerate}

\section{HOD variation with cosmic environment}\label{Sec_HOD}

In the hierarchical structure formation paradigm, galaxies form within dark matter haloes, and the amount of galaxies as well as their properties are primarily governed by the halo mass. This motivates the halo occupation distribution (HOD) framework (see, e.g., \citealp{Benson_2000, Seljak_2000, Peacock_2000, White_2001, Berlind_2002, COORAY_2002}), which describes the average number of galaxies hosted by haloes of a given mass. In its simplest form, the HOD assumes that halo mass alone determines galaxy occupation. In the presence of GAB, galaxy occupation is influenced not only by halo mass but also by other factors, such as the cosmic environment in which haloes reside or halo internal properties such as concentration. More recent HOD extensions introduce decorations upon the basic HOD, allowing them to incorporate GAB within the HOD framework (see e.g., \citealp{Hearin2016}). When there is GAB effect, we can gain valuable insights from the HOD formalism by measuring the occupation for haloes with different secondary halo properties (e.g., \citealp{Artale_2018,Zehavi_2018}). From the last section we showed that GAB can be largely attributed to the cosmic overdensity $\delta_5$. In this section, we further investigate how $\delta_5$ influences the halo–galaxy connection by examining how the HOD varies with $\delta_5$. For completeness, we also analyse how the HOD changes with the cosmic environment as characterised by the smoothed tidal anisotropy $\alpha_5$.

% \begin{figure*}
%     \centering
%     \includegraphics[width=1\linewidth]{figs/HOD_alpha.pdf}
%     \caption{The same as Fig. \ref{HOD_0_delta} but the cosmic environment is now defined by the tidal anisotropy, $\alpha_5$.}
%     \label{HOD_2_delta}
% \end{figure*}
\subsection{HOD variation with overdensity}

We first measure the HOD for both central and satellite galaxies of all haloes, as well as of haloes in the top and bottom quartiles of $\delta_5$ within each halo mass bin. We do the measurements at $z=0$ and $z=2$, which corresponds to the epoch when TNG and SIMBA show the maximum GAB effects, respectively. The results are shown in Fig.~\ref{simba_delta} and Fig.~\ref{tng_delta} for SIMBA and TNG, respectively. At $z=0$, we find that haloes with higher $\delta_5$ are more likely to host central galaxies, with this trend being more pronounced in TNG than in SIMBA. In TNG, haloes with higher $\delta_5$ also tend to host more satellite galaxies, while satellite occupation in SIMBA does not show dependence on the overdensity. The stronger preference for galaxy formation in high-$\delta_5$ haloes in TNG is consistent with the larger total GAB observed in TNG compared to SIMBA at $z=0$. 

At $z=2$, the $\delta_5$-dependence of HOD in TNG is weaker than at $z=0$, in line with the smaller GAB amplitude at $z=2$. For SIMBA, the HOD exhibits variation with $\delta_5$ at slightly higher halo masses than at $z=0$, resulting in a stronger GAB than at $z=0$. The SIMBA and TNG results at $z=2$ are then quite similar, even though they are very different at $z=0$. Notice that in both SIMBA and TNG, the HOD environmental dependence is primarily seen in low-mass haloes. We will further discuss the contribution to the total GAB from low-mass haloes in the next Section.

\subsection{HOD variation with tidal anisotropy}
\begin{figure*}
    \centering
    \includegraphics[width=1\linewidth]{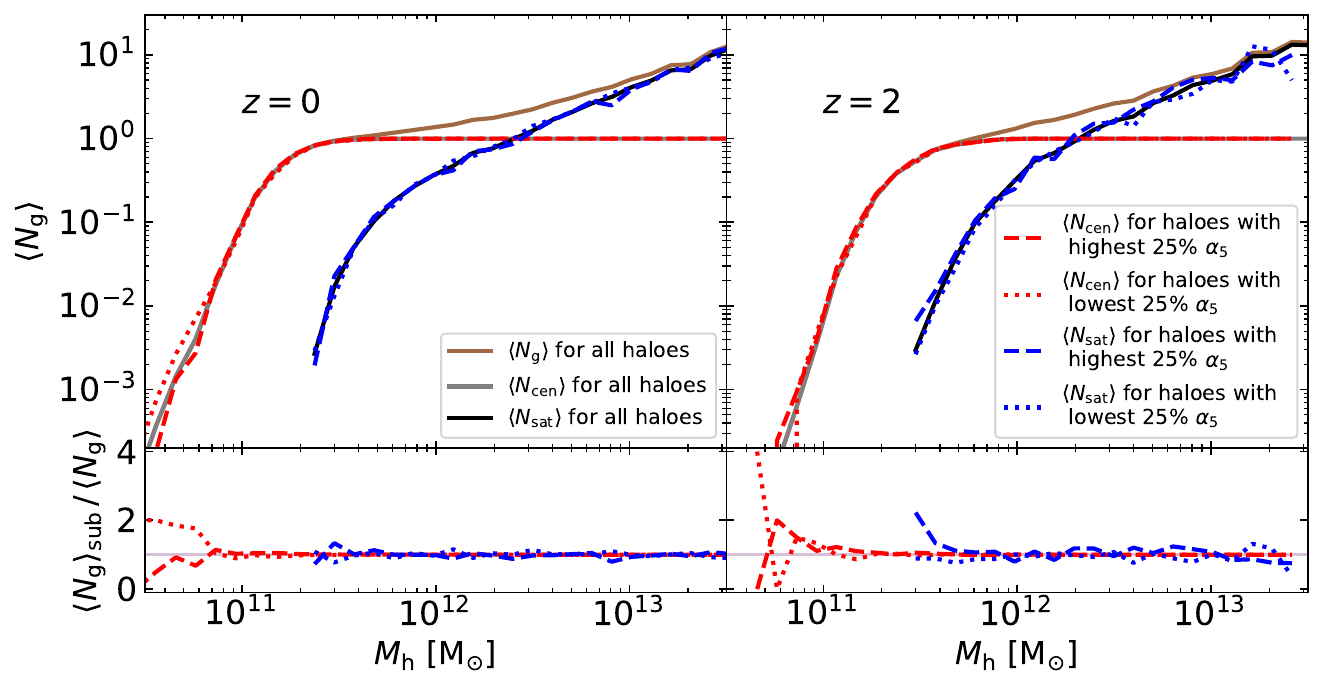}
    \caption{The same as Fig.~\ref{simba_delta}, but with the cosmic environment defined by the smoothed tidal anisotropy $\alpha_5$.}
    \label{simba_alpha}
\end{figure*}

\begin{figure*}
    \centering
    \includegraphics[width=1\linewidth]{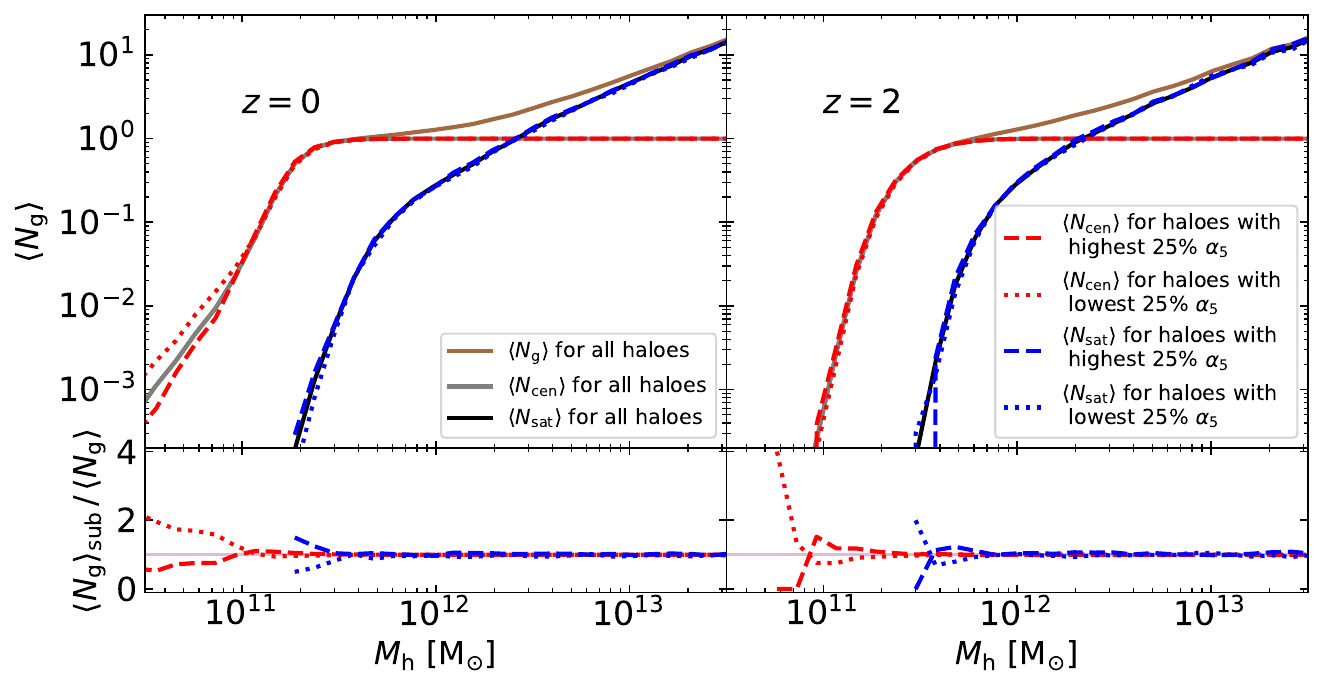}
    \caption{The same as Fig.~\ref{simba_alpha}, but for TNG.}
    \label{tng_alpha}
\end{figure*}
For completeness, we also examine how the HOD varies with the tidal anisotropy environment, $\alpha_5$, in which haloes reside. The results are shown in Fig.~\ref{simba_alpha} and Fig.~\ref{tng_alpha}. At $z=0$, the HOD from both SIMBA and TNG shows very little variation with $\alpha_5$, while at $z=2$ no visible variation with $\alpha_5$ is detected. These results are consistent with the double shuffling with $\alpha_5$ for SIMBA, where we find that GAB has almost no dependence on $\alpha_5$. However, $\alpha_5$ accounts for certain fraction of GAB in TNG, suggesting that $\alpha_5$ affects the galaxy clustering in ways not captured by HOD measurements, such as modulating the spatial distribution of satellite galaxies within haloes. A detailed investigation of this is beyond the scope of the present work and is left for future work.

% In contrast, \citet{Alam_2023} reported a strong HOD dependence on tidal anisotropy, based on galaxy clustering statistics from GAMA \citep{GAMA1,GAMA2} and HOD modelling applied to $N$-body simulations that included $\alpha$-dependent terms. Based on our results, we argue that the apparent dependence found in their study may be driven by the correlation between tidal anisotropy and local overdensity, which was not explicitly removed in their analysis.

\section{GAB from low-mass haloes}\label{Sec_lmh}

In Sec.~\ref{Sec_GAB}, we showed that GAB evolves differently in the two hydrodynamical simulations, and that the cosmic environment as quantified by the smoothed overdensity $\delta_5$ accounts for most of the GAB signal. In Sec.~\ref{Sec_HOD}, we further demonstrated that $\delta_5$ contributes to GAB by modulating the likelihood that low-mass haloes host central and/or satellite galaxies. In this section, we aim to quantify the fraction of the total GAB that originates specifically from low-mass haloes. To this end, we again apply the shuffling technique, but now restrict it to galaxies within either low-mass or high-mass haloes divided by an arbitrary halo mass threshold. The analysis is performed at the redshift where each simulation shows its peak total GAB: $z=2$ for SIMBA and $z=0$ for TNG, so that the contributions from different halo mass regimes are maximally expressed. As a further test of whether GAB arises primarily from haloes in a fixed mass range (which happens to be the low-mass haloes in our baseline sample), or instead comes from haloes that are low-mass relative to the galaxy selection, we raise the stellar mass threshold for galaxy selection in TNG and perform the previous analysis on the new galaxy sample. This is possible thanks to the large simulation volume and thus the high number of galaxies in TNG, which allows us to increase the galaxy stellar mass threshold up to $10^{10}\,\mathrm{M}_\odot$ while maintaining statistical robustness.
\begin{figure}
    \centering
    \includegraphics[width=1\linewidth]{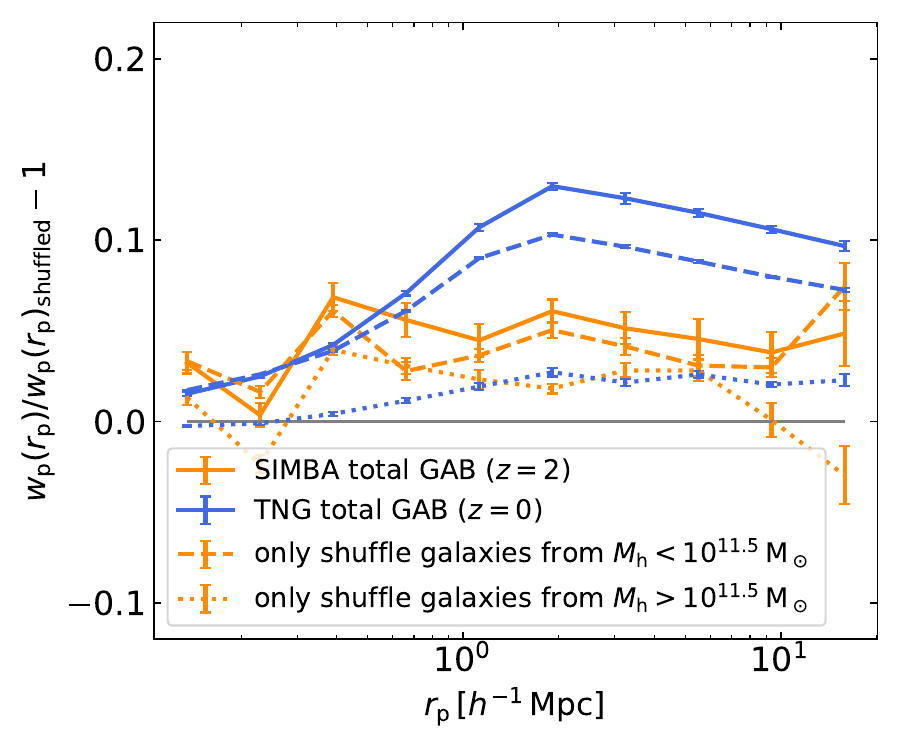}
    \caption{In this figure, we examine the contribution of low-mass and high-mass haloes (defined by a split in total halo mass at $10^{11.5}\,\mathrm{M}_\odot$) to the total galaxy assembly bias (GAB) at the redshift where each simulation exhibits its peak total GAB signal -- $z=2$ for SIMBA and $z=0$ for TNG. The orange (blue) solid line shows the total GAB in SIMBA at $z=2$ (TNG at $z=0$). The dashed (dotted) line shows the GAB measured when only galaxies in haloes with $M_\mathrm{h}<10^{11.5}\,\mathrm{M}_\odot$ ($M_\mathrm{h}>10^{11.5}\,\mathrm{M}_\odot$) are shuffled, for each simulation at the corresponding redshift.}
    \label{fig_lmh}
\end{figure}
\subsection{The baseline sample}

Based on the HOD measurements in Sec.~\ref{Sec_HOD}, we define the dividing mass between low and high halo mass as $M_\mathrm{h} = 10^{11.5}\,\mathrm{M}_\odot$.  The results for the separate shufflings are presented in Fig.~\ref{fig_lmh}. For both SIMBA and TNG, nearly the entire GAB signal originates from low-mass haloes, with a negligible contribution from high-mass haloes. These results align with our findings from the HOD analysis, where we find that the smoothed overdensity $\delta_5$ only affects the occupation for low-mass haloes. This correspondence suggests that one key origin of GAB is that low-mass haloes residing in high-$\delta_5$ environments are more likely to host galaxies.
% At the same time, the significant GAB contribution from high-mass haloes in SIMBA -- absent in TNG -- points to an additional mechanism at play, probably related to baryonic physics, such as differences in feedback implementation between the two simulations.

\begin{figure}
    \centering
    \includegraphics[width=1\linewidth]{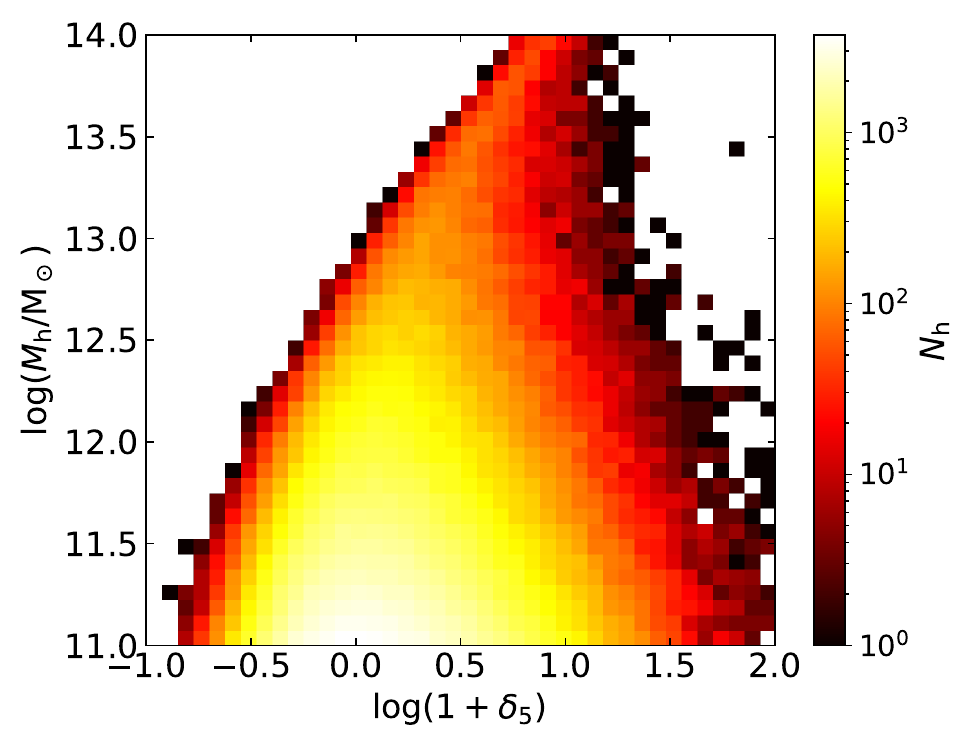}
    
    \caption{Distribution of haloes from TNG at $z=0$ on the $M_\mathrm{h}$–$\delta_5$ plane. The colour map shows the number of haloes in each pixel.}
    \label{shadab}
\end{figure}

This larger GAB effect for low-mass haloes is however perhaps not so surprising, given the statistical relation between halo mass and the overdensity distribution. High-mass haloes are almost always found in highly dense regions, so that there is a larger scatter in $\delta_5$ for low-mass haloes than for high-mass ones. This is illustrated in Fig.~\ref{shadab}, which shows the distribution of haloes from TNG at $z=0$ on the $M_\mathrm{h}$–$\delta_5$ plane. Here, low-mass haloes span nearly three orders of magnitude in $\delta_5$, whereas the most massive haloes barely cover a factor 2 in overdensity.

\subsection{Effect of a higher galaxy stellar mass threshold}
\begin{figure}
    \centering    \includegraphics[width=1\linewidth]{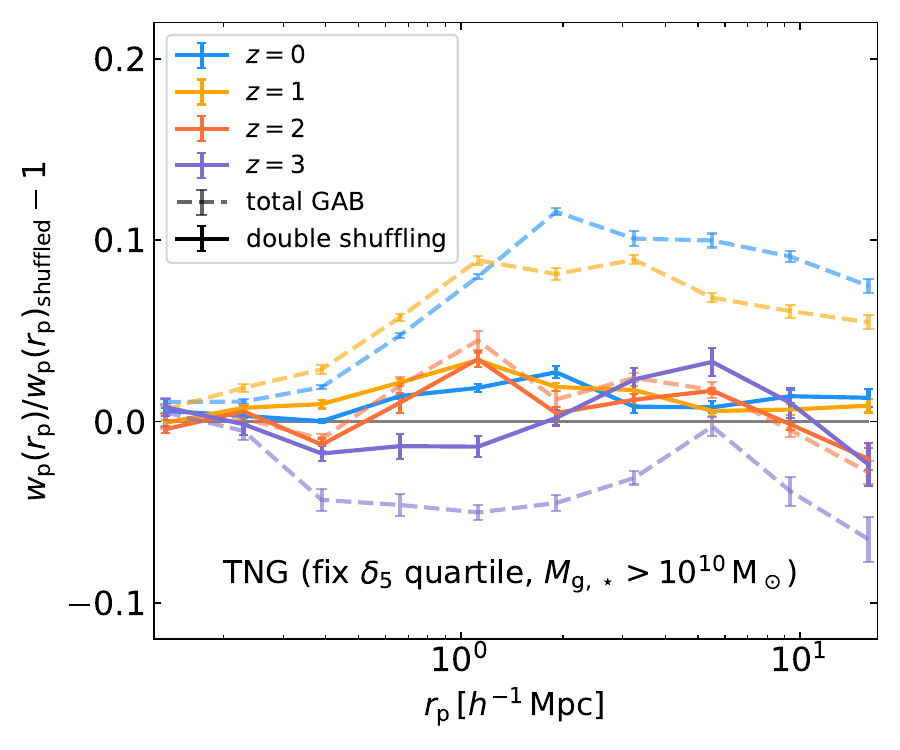}
    \caption{The redshift evolution of the total GAB (dashed lines) and the GAB after removing contribution from the smoothed cosmic overdensity $\delta_5$ (solid lines), using the new TNG galaxy sample selected by applying $M_{\mathrm{g,\star}}>10^{10}\,\mathrm{M_\odot}$.}
    \label{TNG_shuffle_10}
\end{figure}

\begin{figure*}
    \centering    \includegraphics[width=1\linewidth]{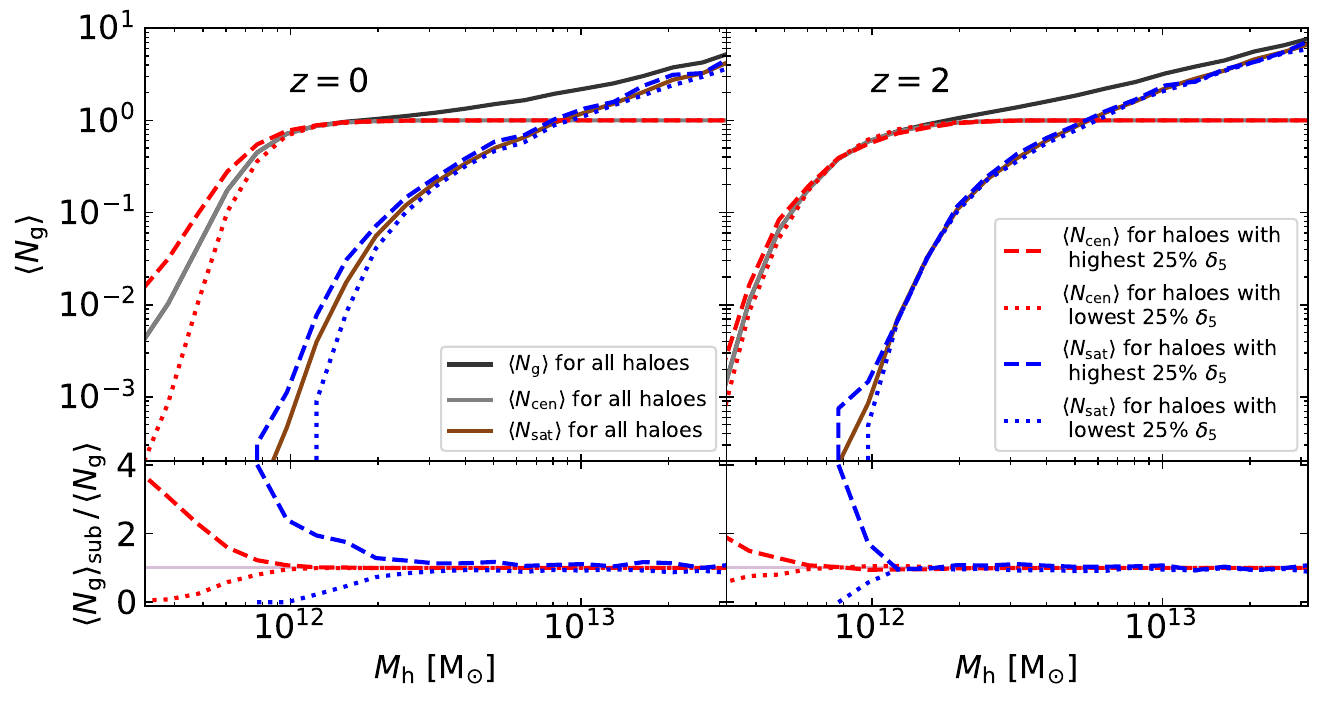}
    \caption{The same as Fig.~\ref{tng_delta}, but using the galaxy sample with stellar mass greater than $10^{10}\,\mathrm{M_\odot}$.}
    \label{HOD_10_tng}
\end{figure*}

\begin{figure}
    \centering    \includegraphics[width=1\linewidth]{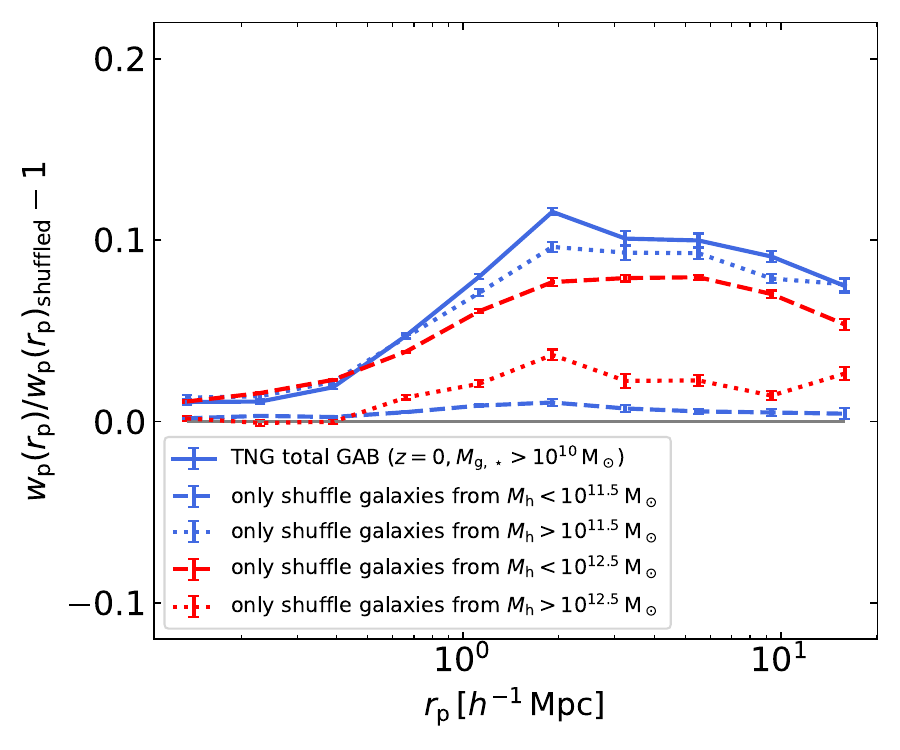}
    \caption{The contribution of low-mass (dashed lines) and high-mass (dotted lines) haloes, defined by the relation between $M_\mathrm{h}$ and $10^{11.5/12.5}\,\mathrm{M}_\odot$, to the total GAB (solid line) in TNG at $z=0$ using the galaxy sample with higher galaxy stellar mass threshold. }
    \label{lmh_10}
\end{figure}
In the previous subsection, we demonstrated that most of the GAB originates from low-mass haloes with $M_\mathrm{h} < 10^{11.5}\,\mathrm{M}_\odot$. Here, we raise the galaxy stellar mass threshold in TNG from $\sim$\,$10^{8.5}\,\mathrm{M}_\odot$ to $10^{10}\,\mathrm{M}_\odot$ in order to test whether GAB arises from a specific halo mass range or is primarily associated with the relatively low-mass haloes capable of hosting galaxies in the sample. The number densities of galaxies corresponding to this higher stellar mass threshold are now $0.011\,h^3\,\mathrm{Mpc}^{-3}$ at $z=0$ and $0.005\,h^3\, \mathrm{Mpc}^{-3}$ at $z=2$.

We repeat the analysis carried out for the baseline sample, examining the redshift evolution of the total GAB signal, the effect of double shuffling with $\delta_5$, and the variation of the HOD with $\delta_5$. The redshift evolution of the total GAB is shown by the dashed lines in Fig.~\ref{TNG_shuffle_10}. In this higher-threshold sample, the total GAB becomes negative at $z=3$, indicating that GAB reduces galaxy clustering at that epoch. As in the baseline case, the strength of GAB increases towards lower redshift, reaching $\sim$\,$10\%$ at $z=0$. Overall, the evolutionary trend resembles that of the baseline sample but with a somewhat weaker effect in general.  

The double-shuffling results, shown by the solid lines in Fig.~\ref{TNG_shuffle_10}, again demonstrate that nearly all of the total GAB can be explained by $\delta_5$, consistent with the baseline analysis. The corresponding HOD variation with $\delta_5$ is presented in Fig.~\ref{HOD_10_tng}. As expected, the HOD shifts toward higher halo masses due to the increased stellar-mass threshold, and $\delta_5$ continues to modulate the probability that low-mass haloes host central and satellite galaxies at $z=0$. At $z=2$, variations of the HOD with $\delta_5$ are not evident, consistent with the fact that the total GAB in the new sample is approximately zero at this redshift.

Finally, Fig.~\ref{lmh_10} shows the results of restricting the shuffling to galaxies within either low-mass or high-mass haloes. Using the previous dividing mass of $M_\mathrm{h} = 10^{11.5}\,\mathrm{M}_\odot$, the GAB signal appears to originate predominantly from high-mass haloes. However, motivated by the HOD variation with $\delta_5$ (Fig.~\ref{HOD_10_tng}), if we instead adopt a higher division at $M_\mathrm{h} = 10^{12.5}\,\mathrm{M}_\odot$, the GAB is found to arise mainly from low-mass haloes.  

In summary, increasing the stellar-mass threshold slightly reduces the total GAB signal while preserving its increasing trend with redshift. The environment parameter $\delta_5$ continues to account for nearly all of the GAB in TNG, acting by modulating the occupancy of low-mass haloes. The shuffling tests further confirm that the dominant contribution to GAB arises from the relatively low-mass haloes that are capable of hosting the selected galaxies.

\section{Fitting HOD variation with overdensity}\label{Sec_fit}

In the preceding sections, we demonstrated that one of the main origins of GAB is that the likelihood for low-mass haloes to host central or satellite galaxies is directly modulated by the cosmic environment in which they reside, as quantified by the smoothed matter overdensity $\delta_5$. Consequently, the HOD is a function of $\delta_5$. Since the HOD remains one of the most widely used approaches for constructing galaxy mocks, incorporating $\delta_5$-dependent deviations into the HOD represents a first step towards generating catalogues that more closely resemble those from hydrodynamical simulations, which currently provide the most physically motivated galaxy catalogues available. In this section,
we therefore provide fitting formulae for the above results, with the aim of describing not only the overall HOD but also its variation with local density.

As regards the form of the overall HOD, we found that the common 5-parameter function introduced by \cite{Zheng2005} did not fit our results so well, especially regarding the relatively gentle onset of non-zero satellite numbers. We therefore propose the following alternative model that provides some additional flexibility:
\begin{equation}
\eqalign{
N_c &= {1\over 1+ (M/M_1)^{-\mu}}; \cr
N_s &= N_c\; {(M/M_2)^\gamma\over 1+ (M/M_3)^{-\nu}}. \cr
}
\label{eq:hodmod}
\end{equation}
In other words, there are characteristic masses $M_1$ and $M_3$ that define the onset of the central and satellite populations, and a characteristic mass $M_2$ at which there is a single satellite. The power-law slopes $\mu$ and $\nu$ determine how rapidly the two galaxy populations `turn on' with halo mass, and they need to be different. Note that, as is common, we require a central galaxy before satellites are allowed to exist. The values of these six free parameters required to fit our various galaxy samples are given in Table \ref{tab:hodpars}. Here, we give results not only for the overall HODs, but also for the samples restricted to the lowest 25\% in $\delta_5$ (Q1) and the highest 25\% (Q4). The parameters tend to be closely coupled, with $M_2/M_1$ close to 10, and $M_3/M_2$ close to 0.4. We also note that $\gamma$ is always close to 1; these may be considered fiducial values, so that the main variation is in $M_1$, $\mu$ and $\nu$.

In examining the fits to the density-dependent data, we were struck by the fact that the main differences with respect to the overall HOD lay almost entirely in the rapidity of onset of the central and satellite populations, as is visually apparent from Figs \ref{simba_delta}, \ref{tng_delta}, and \ref{HOD_10_tng}. We therefore carried out a fit to the data in the $\delta_5$ quartiles in which only the slopes $\mu$ and $\nu$ were allowed to vary, with all other parameters being fixed at the values required by the mean HOD. This provides a convenient way of quantifying the density dependence, as follows. We consider the rank of the $\delta_5$ values, $r_\delta$, which varies from 0 at the lowest density to 1 at the maximum; we assume that the lower and upper quartiles give us results for ranks 0.125 and 0.875. We then assume that the key slopes in the HOD depend linearly on the rank:
\begin{equation}
    \mu = \mu_{\rm tot} -C_\delta\, (r_\delta - 0.5),
\end{equation}
and similarly for $\nu$. Here, $\mu_{\rm tot}$ is the slope appropriate for the total HOD. The sign is chosen so that a positive value of $C_\delta$ means that $\mu$ declines at high densities, whereas the onset of galaxies becomes more rapid at low densities. The density dependence of central galaxies is generally stronger than for satellites, with $C_\delta(\mu) \sim 2 C_\delta(\nu)$. Indeed, for some subsets (particularly SIMBA) we see no evidence for a density dependence of satellite onset, even though there is an effect on centrals.

The differences between SIMBA and TNG are also marked in terms of the behaviour at different redshifts. TNG has a marked density sensitivity at both redshift 0 and 2, whereas SIMBA shows less of an effect at redshift 2. TNG does show a reduction in density dependence at redshift 2 for the more massive galaxy selection, but we are unable to supply a comparison with SIMBA in this case. There is therefore no clear-cut prediction from these simulations about the degree of density dependence of the HOD that might be seen in practice. However, these results do suggest that it would be interesting to search for variations of the above form, with $C_\delta$ of order unity. Similarly, more realistic mock data could be produced by considering a range of values at this level. This would certainly be better than the unmotivated assumption of zero variation with density, and marginalising over the unknown $C_\delta$ values could yield more robust uncertainties on cosmological parameters of interest.

\begin{table}
\caption{Best-fit HOD parameters for the different samples. The symbol $M$ denotes $\log_{10}$ of mass in units of $\msun$. The minimum stellar mass defining the galaxy samples studied at redshift $z$ is denoted by $M_*$. The data for the lowest and highest quartiles of $\delta_5$ density (Q1 and Q4) are described by varying only the cutoff slopes, $\mu$ and $\nu$. The density coefficients assume a linear relation of these slopes to rank in $\delta_5$. For some cases, there is no evidence for a change in the numbers of satellites with $\delta_5$, and a slope is not quoted.} 
\label{tab:hodpars}
\begin{tabular}{lcccccccc}

\hline
Sample & $z$ & $M_*$ & $M_1$ & $M_2$ & $M_3$ & $\gamma$ & $\mu$ & $\nu$ \\
\hline

TNG & 0 & 8.7 & 11.29 & 12.39 & 11.83 & 1.01 & 5.06 & 2.81 \\
Q1 & \multispan6 & 8.58 & 4.14 \\
Q4 & \multispan6 & 3.39 & 2.25 \\
$\delta$ coeff & \multispan6 & 6.92 & 2.52 \\
\hline

TNG & 0 & 10 & 11.92 & 12.92 & 12.44 & 1.01 & 5.31 & 4.10 \\
Q1 & \multispan6 & 7.95 & 5.11 \\
Q4 & \multispan6 & 4.11 & 3.58 \\
$\delta$ coeff & \multispan6 & 5.12 & 2.04 \\
\hline

TNG & 2 & 9.1 & 11.45 & 12.30 & 11.94 & 0.99 & 5.47 & 3.93 \\
Q1 & \multispan6 & 6.20 & 4.43 \\
Q4 & \multispan6 & 4.60 & 3.23 \\
$\delta$ coeff & \multispan6 & 2.13 & 1.60 \\
\hline

TNG & 2 & 10 & 11.93 & 12.70 & 12.44 & 0.98 & 5.06 & 3.90 \\
Q1 & \multispan6 & 5.52 & 4.00 \\
Q4 & \multispan6 & 4.71 & 3.95 \\
$\delta$ coeff & \multispan6 & 1.08 & -- \\
\hline

SIMBA & 0 & 8.8 & 11.18 & 12.39 & 11.69 & 0.95 & 5.51 & 3.77 \\
Q1 & \multispan6 & 5.99 & 4.22 \\
Q4 & \multispan6 & 5.14 & 4.21 \\
$\delta$ coeff & \multispan6 & 1.13 & -- \\
\hline

SIMBA & 2 & 8.8 & 11.43 & 12.26 & 11.94 & 0.90 & 4.26 & 3.28 \\
Q1 & \multispan6 & 4.83 & 3.31 \\
Q4 & \multispan6 & 3.82 & 2.86 \\
$\delta$ coeff & \multispan6 & 1.35 & -- \\
\hline
\end{tabular}
    
\end{table}

\section{Summary and Conclusions}\label{conclusion}

We have studied the strength and redshift evolution of galaxy assembly bias (GAB) seen in mock galaxy catalogues selected by stellar mass, using two state-of-the-art cosmological hydrodynamical simulations: SIMBA and \textsc{Illustris}TNG. In particular, we investigated whether GAB is connected to the cosmic environment in which haloes reside, as quantified by the smoothed overdensity $\delta_5$ and the tidal anisotropy $\alpha_5$. This was explored using two complementary approaches: the double-shuffling technique and the analysis of HOD variations across different environments. Additionally, we examined the contribution of low-mass and high-mass haloes to the total GAB signal by applying shuffling only within each mass regime. We also repeated the analysis with a higher selection threshold in galaxy stellar mass, in order to investigate how our conclusions change with the selection criteria. Our main findings are summarised below:
\begin{enumerate}
\item SIMBA and TNG exhibit different GAB evolution trends, as measured by the ratio of the projected correlation functions of the original and shuffled samples. In both simulations, GAB is negligible at $z=5$ and increases to $\sim$\,$5\%$ by $z=2$. After that, GAB in SIMBA declines steadily, reaching nearly zero at $z=0$, while in TNG it continues to grow, reaching $\sim$\,$10\%$ at $z=0$.

\item Using the double-shuffling technique, we find that most of the GAB seen in SIMBA and TNG can be attributed to the modulation of the halo–galaxy connection by $\delta_5$. 
%\sout{but in SIMBA, $\delta_5$ accounts for approximately half of the total GAB. }
In both simulations, the contribution from $\alpha_5$ is negligible (SIMBA) or minor (TNG).

\item The HOD measured in both simulations shows a dependence on $\delta_5$ for low-mass haloes, consistent with the findings that  $\delta_5$ can account for most of the GAB. No significant HOD dependence on $\alpha_5$ is detected in either simulations, suggesting the minor GAB dependence on $\alpha_5$ in TNG could originate from mechanisms not captured by HOD measurements, such as the satellite galaxy distribution within haloes.

\item A mass-split shuffling analysis shows that most of the GAB in both SIMBA and TNG originates from low-mass haloes ($M_\mathrm{h}<10^{11.5}\,\mathrm{M}_\odot$). This is consistent with the fraction of the GAB accounted for by $\delta_5$ in the simulations, suggesting that a primary origin of GAB is the enhanced likelihood for low-mass haloes in high-$\delta_5$ regions to host galaxies.

\item We increased the selection threshold in galaxy stellar mass from $\simeq 10^{8.5}\msun$ to $10^{10}\msun$ (only possible in TNG, on account of its larger volume). This change
slightly reduces the total GAB signal while preserving its increasing trend with redshift. The environment parameter $\delta_5$ continues to account for nearly all of the GAB in TNG, acting by modulating the occupancy of low-mass haloes. The shuffling tests confirm that the dominant contribution to GAB arises from relatively low-mass haloes, near to the `turn-on' mass in the HOD.

\item We provide a fitting formulation that describes how the HOD deviates from the average in different cosmic environment, as quantified by $\delta_5$, based on measurements from SIMBA and TNG. This formulation can be applied to generate HOD-based galaxy mocks that more closely resemble those produced by hydrodynamical simulations. 

\end{enumerate}

Our results show that GAB can alter galaxy clustering by up to $\sim$\,$10\%$ in amplitude. Accurate modelling of GAB is therefore essential for the correct interpretation of the precise data being delivered by modern galaxy redshift surveys. By comparing the redshift evolution of GAB in two state-of-the-art cosmological hydrodynamical simulations, SIMBA and \textsc{Illustris}TNG, we found significantly different trends between them. This difference highlights the strong dependence of GAB on the baryonic physics implemented in simulations, potentially providing a route towards understanding its physical origin. 

Our results suggest that the dominant contributing factor to GAB is expected to be the enhanced likelihood for low-mass haloes in high-density regions to host galaxies. Taken together with CAMELS results on the differing baryonic physics of SIMBA and TNG, our findings suggest that low-mass haloes in high-overdensity environments are more likely to form galaxies; however, this tendency can be suppressed by different recipes for AGN feedback from nearby massive haloes.
In comparison, we find that modifications of galaxy populations by the local tidal anisotropy are expected to be sub-dominant to the effect of overdensity. This prediction is in conflict with some claims based on observations \citep{Alam_2023}; further studies of this discrepancy will be important in order to establish with certainty the empirical degree of assembly bias, and to gain a better understanding of its origin.

%Identifying and characterising the physical mechanism behind this phenomenon will be important for a comprehensive understanding of GAB, which we aim to study in the future works.
% However, further investigation is required to confirm this scenario and to quantify its role across different environments and redshifts.
% Additionally, we find that in SIMBA, roughly half of the total GAB at $z=2$ arises from high-mass haloes, yet this component does not appear to be explained by $\delta_5$. This suggests the existence of a separate mechanism -- possibly tied to baryonic physics -- that drives GAB in the high-mass regime. 
%Overall, our work represents a step toward uncovering the physical processes responsible for galaxy assembly bias, and underscores the importance of including baryonic effects in models of galaxy clustering.

\section*{Acknowledgements}

HGY is grateful for support from the China Scholarship Council (CSC) under Grant No. CSC 202306340042. We are grateful to Shadab Alam and to Sergio Contreras for helpful comments that allowed us to tighten some of our arguments. We thank the anonymous referee for their constructive comments, which helped us improve the discussion of our results.

%%%%%%%%%%%%%%%%%%%%%%%%%%%%%%%%%%%%%%%%%%%%%%%%%%
\section*{Data Availability}

The data of SIMBA simulation suite is publicly accessible from their websites: \href{http://simba.roe.ac.uk}{\url{http://simba.roe.ac.uk}} \citep{SIMBA}. The \textsc{Illustris}TNG simulations, including TNG300-1, are publicly available and
accessible at \href{https://www.tng-project.org}{\url{https://www.tng-project.org}} \citep{TNG_DR}.

\bibliographystyle{mnras}
\bibliography{mybib} 

@ARTICLE{Zheng2005,
       author = {{Zheng}, Zheng and {Berlind}, Andreas A. and {Weinberg}, David H. and {Benson}, Andrew J. and {Baugh}, Carlton M. and {Cole}, Shaun and {Dav{\'e}}, Romeel and {Frenk}, Carlos S. and {Katz}, Neal and {Lacey}, Cedric G.},
        title = "{Theoretical Models of the Halo Occupation Distribution: Separating Central and Satellite Galaxies}",
      journal = {\apj},
         year = 2005,
        month = nov,
       volume = {633},
       number = {2},
        pages = {791-809},
          doi = {10.1086/466510},
archivePrefix = {arXiv},
       eprint = {astro-ph/0408564},
 primaryClass = {astro-ph},
       adsurl = {https://ui.adsabs.harvard.edu/abs/2005ApJ...633..791Z}
}

@ARTICLE{Paranjape2018,
       author = {{Paranjape}, Aseem and {Hahn}, Oliver and {Sheth}, Ravi K.},
        title = "{Halo assembly bias and the tidal anisotropy of the local halo environment}",
      journal = {\mnras},
         year = 2018,
        month = may,
       volume = {476},
       number = {3},
        pages = {3631-3647},
          doi = {10.1093/mnras/sty496},
archivePrefix = {arXiv},
       eprint = {1706.09906},
 primaryClass = {astro-ph.CO},
       adsurl = {https://ui.adsabs.harvard.edu/abs/2018MNRAS.476.3631P}
}

@ARTICLE{SIMBA,
       author = {{Dav{\'e}}, Romeel and {Angl{\'e}s-Alc{\'a}zar}, Daniel and {Narayanan}, Desika and {Li}, Qi and {Rafieferantsoa}, Mika H. and {Appleby}, Sarah},
        title = "{SIMBA: Cosmological simulations with black hole growth and feedback}",
      journal = {\mnras},
         year = 2019,
        month = jun,
       volume = {486},
       number = {2},
        pages = {2827-2849},
          doi = {10.1093/mnras/stz937},
archivePrefix = {arXiv},
       eprint = {1901.10203},
 primaryClass = {astro-ph.GA},
       adsurl = {https://ui.adsabs.harvard.edu/abs/2019MNRAS.486.2827D}
}

@article{Planck2015,
   title={Planck2015 results: XIII. Cosmological parameters},
   volume={594},
   ISSN={1432-0746},
   url={http://dx.doi.org/10.1051/0004-6361/201525830},
   DOI={10.1051/0004-6361/201525830},
   journal={Astronomy \& Astrophysics},
   publisher={EDP Sciences},
   author={Ade, P. A. R. and Aghanim, N. and Arnaud, M. and Ashdown, M. and Aumont, J. and Baccigalupi, C. and Banday, A. J. and Barreiro, R. B. and Bartlett, J. G. and Bartolo, N. and Battaner, E. and Battye, R. and Benabed, K. and Benoît, A. and Benoit-Lévy, A. and Bernard, J.-P. and Bersanelli, M. and Bielewicz, P. and Bock, J. J. and Bonaldi, A. and Bonavera, L. and Bond, J. R. and Borrill, J. and Bouchet, F. R. and Boulanger, F. and Bucher, M. and Burigana, C. and Butler, R. C. and Calabrese, E. and Cardoso, J.-F. and Catalano, A. and Challinor, A. and Chamballu, A. and Chary, R.-R. and Chiang, H. C. and Chluba, J. and Christensen, P. R. and Church, S. and Clements, D. L. and Colombi, S. and Colombo, L. P. L. and Combet, C. and Coulais, A. and Crill, B. P. and Curto, A. and Cuttaia, F. and Danese, L. and Davies, R. D. and Davis, R. J. and de Bernardis, P. and de Rosa, A. and de Zotti, G. and Delabrouille, J. and Désert, F.-X. and Di Valentino, E. and Dickinson, C. and Diego, J. M. and Dolag, K. and Dole, H. and Donzelli, S. and Doré, O. and Douspis, M. and Ducout, A. and Dunkley, J. and Dupac, X. and Efstathiou, G. and Elsner, F. and Enßlin, T. A. and Eriksen, H. K. and Farhang, M. and Fergusson, J. and Finelli, F. and Forni, O. and Frailis, M. and Fraisse, A. A. and Franceschi, E. and Frejsel, A. and Galeotta, S. and Galli, S. and Ganga, K. and Gauthier, C. and Gerbino, M. and Ghosh, T. and Giard, M. and Giraud-Héraud, Y. and Giusarma, E. and Gjerløw, E. and González-Nuevo, J. and Górski, K. M. and Gratton, S. and Gregorio, A. and Gruppuso, A. and Gudmundsson, J. E. and Hamann, J. and Hansen, F. K. and Hanson, D. and Harrison, D. L. and Helou, G. and Henrot-Versillé, S. and Hernández-Monteagudo, C. and Herranz, D. and Hildebrandt, S. R. and Hivon, E. and Hobson, M. and Holmes, W. A. and Hornstrup, A. and Hovest, W. and Huang, Z. and Huffenberger, K. M. and Hurier, G. and Jaffe, A. H. and Jaffe, T. R. and Jones, W. C. and Juvela, M. and Keihänen, E. and Keskitalo, R. and Kisner, T. S. and Kneissl, R. and Knoche, J. and Knox, L. and Kunz, M. and Kurki-Suonio, H. and Lagache, G. and Lähteenmäki, A. and Lamarre, J.-M. and Lasenby, A. and Lattanzi, M. and Lawrence, C. R. and Leahy, J. P. and Leonardi, R. and Lesgourgues, J. and Levrier, F. and Lewis, A. and Liguori, M. and Lilje, P. B. and Linden-Vørnle, M. and López-Caniego, M. and Lubin, P. M. and Macías-Pérez, J. F. and Maggio, G. and Maino, D. and Mandolesi, N. and Mangilli, A. and Marchini, A. and Maris, M. and Martin, P. G. and Martinelli, M. and Martínez-González, E. and Masi, S. and Matarrese, S. and McGehee, P. and Meinhold, P. R. and Melchiorri, A. and Melin, J.-B. and Mendes, L. and Mennella, A. and Migliaccio, M. and Millea, M. and Mitra, S. and Miville-Deschênes, M.-A. and Moneti, A. and Montier, L. and Morgante, G. and Mortlock, D. and Moss, A. and Munshi, D. and Murphy, J. A. and Naselsky, P. and Nati, F. and Natoli, P. and Netterfield, C. B. and Nørgaard-Nielsen, H. U. and Noviello, F. and Novikov, D. and Novikov, I. and Oxborrow, C. A. and Paci, F. and Pagano, L. and Pajot, F. and Paladini, R. and Paoletti, D. and Partridge, B. and Pasian, F. and Patanchon, G. and Pearson, T. J. and Perdereau, O. and Perotto, L. and Perrotta, F. and Pettorino, V. and Piacentini, F. and Piat, M. and Pierpaoli, E. and Pietrobon, D. and Plaszczynski, S. and Pointecouteau, E. and Polenta, G. and Popa, L. and Pratt, G. W. and Prézeau, G. and Prunet, S. and Puget, J.-L. and Rachen, J. P. and Reach, W. T. and Rebolo, R. and Reinecke, M. and Remazeilles, M. and Renault, C. and Renzi, A. and Ristorcelli, I. and Rocha, G. and Rosset, C. and Rossetti, M. and Roudier, G. and Rouillé d’Orfeuil, B. and Rowan-Robinson, M. and Rubiño-Martín, J. A. and Rusholme, B. and Said, N. and Salvatelli, V. and Salvati, L. and Sandri, M. and Santos, D. and Savelainen, M. and Savini, G. and Scott, D. and Seiffert, M. D. and Serra, P. and Shellard, E. P. S. and Spencer, L. D. and Spinelli, M. and Stolyarov, V. and Stompor, R. and Sudiwala, R. and Sunyaev, R. and Sutton, D. and Suur-Uski, A.-S. and Sygnet, J.-F. and Tauber, J. A. and Terenzi, L. and Toffolatti, L. and Tomasi, M. and Tristram, M. and Trombetti, T. and Tucci, M. and Tuovinen, J. and Türler, M. and Umana, G. and Valenziano, L. and Valiviita, J. and Van Tent, F. and Vielva, P. and Villa, F. and Wade, L. A. and Wandelt, B. D. and Wehus, I. K. and White, M. and White, S. D. M. and Wilkinson, A. and Yvon, D. and Zacchei, A. and Zonca, A.},
   year={2016},
   month=sep, pages={A13} }

@ARTICLE{TNG_DR,
       author = {{Nelson}, Dylan and {Springel}, Volker and {Pillepich}, Annalisa and {Rodriguez-Gomez}, Vicente and {Torrey}, Paul and {Genel}, Shy and {Vogelsberger}, Mark and {Pakmor}, Ruediger and {Marinacci}, Federico and {Weinberger}, Rainer and {Kelley}, Luke and {Lovell}, Mark and {Diemer}, Benedikt and {Hernquist}, Lars},
        title = "{The IllustrisTNG simulations: public data release}",
      journal = {Computational Astrophysics and Cosmology},
         year = 2019,
        month = may,
       volume = {6},
       number = {1},
          eid = {2},
        pages = {2},
          doi = {10.1186/s40668-019-0028-x},
archivePrefix = {arXiv},
       eprint = {1812.05609},
 primaryClass = {astro-ph.GA},
       adsurl = {https://ui.adsabs.harvard.edu/abs/2019ComAC...6....2N}
}

@article{GIZMO,
   title={A new class of accurate, mesh-free hydrodynamic simulation methods},
   volume={450},
   ISSN={1365-2966},
   url={http://dx.doi.org/10.1093/mnras/stv195},
   DOI={10.1093/mnras/stv195},
   number={1},
   journal={Monthly Notices of the Royal Astronomical Society},
   publisher={Oxford University Press (OUP)},
   author={Hopkins, Philip F.},
   year={2015},
   month=apr, pages={53–110} }

@ARTICLE{Hopkins_2017,
       author = {{Hopkins}, Philip F.},
        title = "{A New Public Release of the GIZMO Code}",
      journal = {arXiv:1712.01294},
         year = 2017,
        month = dec,
          doi = {10.48550/arXiv.1712.01294}
}

@article{AREPO,
   title={E pur si muove:Galilean-invariant cosmological hydrodynamical simulations on a moving mesh},
   volume={401},
   ISSN={1365-2966},
   url={http://dx.doi.org/10.1111/j.1365-2966.2009.15715.x},
   DOI={10.1111/j.1365-2966.2009.15715.x},
   number={2},
   journal={Monthly Notices of the Royal Astronomical Society},
   publisher={Oxford University Press (OUP)},
   author={Springel, Volker},
   year={2010},
   month=jan, pages={791–851} }

@article{Subfind,
   title={Populating a cluster of galaxies - I. Results at \fontshape{it}{z}=0},
   volume={328},
   ISSN={1365-2966},
   url={http://dx.doi.org/10.1046/j.1365-8711.2001.04912.x},
   DOI={10.1046/j.1365-8711.2001.04912.x},
   number={3},
   journal={Monthly Notices of the Royal Astronomical Society},
   publisher={Oxford University Press (OUP)},
   author={Springel, Volker and White, Simon D. M. and Tormen, Giuseppe and Kauffmann, Guinevere},
   year={2001},
   month=dec, pages={726–750} }

@article{Croton_2007,
   title={Halo assembly bias and its effects on galaxy clustering},
   volume={374},
   ISSN={1365-2966},
   url={http://dx.doi.org/10.1111/j.1365-2966.2006.11230.x},
   DOI={10.1111/j.1365-2966.2006.11230.x},
   number={4},
   journal={Monthly Notices of the Royal Astronomical Society},
   publisher={Oxford University Press (OUP)},
   author={Croton, D. J. and Gao, L. and White, S. D. M.},
   year={2007},
   month=feb, pages={1303–1309} }

@article{Contreras_2019,
   title={The evolution of assembly bias},
   volume={484},
   ISSN={1365-2966},
   url={http://dx.doi.org/10.1093/mnras/stz018},
   DOI={10.1093/mnras/stz018},
   number={1},
   journal={Monthly Notices of the Royal Astronomical Society},
   publisher={Oxford University Press (OUP)},
   author={Contreras, S and Zehavi, I and Padilla, N and Baugh, C M and Jiménez, E and Lacerna, I},
   year={2019},
   month=jan, pages={1133–1148} }

@article{halotools,
   title={Forward Modeling of Large-scale Structure: An Open-source Approach with Halotools},
   volume={154},
   ISSN={1538-3881},
   url={http://dx.doi.org/10.3847/1538-3881/aa859f},
   DOI={10.3847/1538-3881/aa859f},
   number={5},
   journal={The Astronomical Journal},
   publisher={American Astronomical Society},
   author={Hearin, Andrew P. and Campbell, Duncan and Tollerud, Erik and Behroozi, Peter and Diemer, Benedikt and Goldbaum, Nathan J. and Jennings, Elise and Leauthaud, Alexie and Mao, Yao-Yuan and More, Surhud and Parejko, John and Sinha, Manodeep and Sipöcz, Brigitta and Zentner, Andrew},
   year={2017},
   month=oct, pages={190} }

@ARTICLE{Hearin2016,
       author = {{Hearin}, Andrew P. and {Zentner}, Andrew R. and {van den Bosch}, Frank C. and {Campbell}, Duncan and {Tollerud}, Erik},
        title = "{Introducing decorated HODs: modelling assembly bias in the galaxy-halo connection}",
      journal = {\mnras},
         year = 2016,
        month = aug,
       volume = {460},
       number = {3},
        pages = {2552-2570},
          doi = {10.1093/mnras/stw840},
archivePrefix = {arXiv},
       eprint = {1512.03050},
 primaryClass = {astro-ph.CO},
       adsurl = {https://ui.adsabs.harvard.edu/abs/2016MNRAS.460.2552H}
}

@article{Benson_2000,
   title={The nature of galaxy bias and clustering},
   volume={311},
   ISSN={1365-2966},
   url={http://dx.doi.org/10.1046/j.1365-8711.2000.03101.x},
   DOI={10.1046/j.1365-8711.2000.03101.x},
   number={4},
   journal={Monthly Notices of the Royal Astronomical Society},
   publisher={Oxford University Press (OUP)},
   author={Benson, A. J. and Cole, S. and Frenk, C. S. and Baugh, C. M. and Lacey, C. G.},
   year={2000},
   month=feb, pages={793–808} }

@article{Seljak_2000,
   title={Analytic model for galaxy and dark matter clustering},
   volume={318},
   ISSN={1365-2966},
   url={http://dx.doi.org/10.1046/j.1365-8711.2000.03715.x},
   DOI={10.1046/j.1365-8711.2000.03715.x},
   number={1},
   journal={Monthly Notices of the Royal Astronomical Society},
   publisher={Oxford University Press (OUP)},
   author={Seljak, U.},
   year={2000},
   month=oct, pages={203–213} }

@article{Peacock_2000,
   title={Halo occupation numbers and galaxy bias},
   volume={318},
   ISSN={1365-2966},
   url={http://dx.doi.org/10.1046/j.1365-8711.2000.03779.x},
   DOI={10.1046/j.1365-8711.2000.03779.x},
   number={4},
   journal={Monthly Notices of the Royal Astronomical Society},
   publisher={Oxford University Press (OUP)},
   author={Peacock, J. A. and Smith, R. E.},
   year={2000},
   month=nov, pages={1144–1156} }

@article{White_2001,
   title={The Halo Model and Numerical Simulations},
   volume={550},
   ISSN={0004-637X},
   url={http://dx.doi.org/10.1086/319644},
   DOI={10.1086/319644},
   number={2},
   journal={The Astrophysical Journal},
   publisher={American Astronomical Society},
   author={White, Martin and Hernquist, Lars and Springel, Volker},
   year={2001},
   month=apr, pages={L129–L132} }

@ARTICLE{Berlind_2002,
       author = {{Berlind}, Andreas A. and {Weinberg}, David H.},
        title = "{The Halo Occupation Distribution: Toward an Empirical Determination of the Relation between Galaxies and Mass}",
      journal = {\apj},
         year = 2002,
        month = aug,
       volume = {575},
       number = {2},
        pages = {587-616},
          doi = {10.1086/341469},
archivePrefix = {arXiv},
       eprint = {astro-ph/0109001},
 primaryClass = {astro-ph},
       adsurl = {https://ui.adsabs.harvard.edu/abs/2002ApJ...575..587B}
}

@article{COORAY_2002,
   title={Halo models of large scale structure},
   volume={372},
   ISSN={0370-1573},
   url={http://dx.doi.org/10.1016/S0370-1573(02)00276-4},
   DOI={10.1016/s0370-1573(02)00276-4},
   number={1},
   journal={Physics Reports},
   publisher={Elsevier BV},
   author={Cooray, A and Sheth, R},
   year={2002},
   month=dec, pages={1–129} }

@article{Alam_2023,
   title={Impact of tidal environment on galaxy clustering in GAMA},
   volume={527},
   ISSN={1365-2966},
   url={http://dx.doi.org/10.1093/mnras/stad3423},
   DOI={10.1093/mnras/stad3423},
   number={2},
   journal={Monthly Notices of the Royal Astronomical Society},
   publisher={Oxford University Press (OUP)},
   author={Alam, Shadab and Paranjape, Aseem and Peacock, John A},
   year={2023},
   month=nov, pages={3771–3787} }

@article{Wang_2024,
   title={The beyond-halo mass effects of the cosmic web environment on galaxies},
   volume={532},
   ISSN={1365-2966},
   url={http://dx.doi.org/10.1093/mnras/stae1805},
   DOI={10.1093/mnras/stae1805},
   number={4},
   journal={Monthly Notices of the Royal Astronomical Society},
   publisher={Oxford University Press (OUP)},
   author={Wang, Kuan and Avestruz, Camille and Guo, Hong and Wang, Wei and Wang, Peng},
   year={2024},
   month=jul, pages={4616–4631} }

@ARTICLE{Guo_2013,
       author = {{Guo}, Qi and {White}, Simon and {Angulo}, Raul E. and {Henriques}, Bruno and {Lemson}, Gerard and {Boylan-Kolchin}, Michael and {Thomas}, Peter and {Short}, Chris},
        title = "{Galaxy formation in WMAP1 and WMAP7 cosmologies}",
      journal = {\mnras},
         year = 2013,
        month = jan,
       volume = {428},
       number = {2},
        pages = {1351-1365},
          doi = {10.1093/mnras/sts115},
archivePrefix = {arXiv},
       eprint = {1206.0052},
 primaryClass = {astro-ph.CO},
       adsurl = {https://ui.adsabs.harvard.edu/abs/2013MNRAS.428.1351G}
}

@article{Hadzhiyska_2020,
   title={Limitations to the ‘basic’ HOD model and beyond},
   volume={493},
   ISSN={1365-2966},
   url={http://dx.doi.org/10.1093/mnras/staa623},
   DOI={10.1093/mnras/staa623},
   number={4},
   journal={Monthly Notices of the Royal Astronomical Society},
   publisher={Oxford University Press (OUP)},
   author={Hadzhiyska, Boryana and Bose, Sownak and Eisenstein, Daniel and Hernquist, Lars and Spergel, David N},
   year={2020},
   month=mar, pages={5506–5519} }

@article{Xu_2021,
   title={Dissecting and modelling galaxy assembly bias},
   volume={502},
   ISSN={1365-2966},
   url={http://dx.doi.org/10.1093/mnras/stab100},
   DOI={10.1093/mnras/stab100},
   number={3},
   journal={Monthly Notices of the Royal Astronomical Society},
   publisher={Oxford University Press (OUP)},
   author={Xu, Xiaoju and Zehavi, Idit and Contreras, Sergio},
   year={2021},
   month=jan, pages={3242–3263} }

@ARTICLE{Moreno_2025,
      author={Sergio García-Moreno and Jonás Chaves-Montero},
      title="{Measuring and predicting galaxy assembly bias across galaxy samples}", 
      year={2025},
      journal={arXiv:2504.06770},
      doi={10.48550/arXiv.2504.06770} 
}

@article{Wechsler_2018,
   title={The Connection Between Galaxies and Their Dark Matter Halos},
   volume={56},
   ISSN={1545-4282},
   url={http://dx.doi.org/10.1146/annurev-astro-081817-051756},
   DOI={10.1146/annurev-astro-081817-051756},
   number={1},
   journal={Annual Review of Astronomy and Astrophysics},
   publisher={Annual Reviews},
   author={Wechsler, Risa H. and Tinker, Jeremy L.},
   year={2018},
   month=sep, pages={435–487} }

@article{CAMELS_Ni,
      title="{The CAMELS project: Expanding the galaxy formation model space with new ASTRID and 28-parameter TNG and SIMBA suites}", 
      author={Yueying Ni and Shy Genel and Daniel Anglés-Alcázar and Francisco Villaescusa-Navarro and Yongseok Jo and Simeon Bird and Tiziana Di Matteo and Rupert Croft and Nianyi Chen and Natalí S. M. de Santi and Matthew Gebhardt and Helen Shao and Shivam Pandey and Lars Hernquist and Romeel Dave},
      year={2023},
      journal={arXiv:2304.02096},
      archivePrefix={arXiv},
      primaryClass={astro-ph.CO},
      url={https://arxiv.org/abs/2304.02096}, 
}

@ARTICLE{DESI_overview,
       author = {{DESI Collaboration} and {Adame}, A.~G. and {Aguilar}, J. and {Ahlen}, S. and {Alam}, S. and {Aldering}, G. and {Alexander}, D.~M. and {Alfarsy}, R. and {Allende Prieto}, C. and {Alvarez}, M. and {Alves}, O. and {Anand}, A. and {Andrade-Oliveira}, F. and {Armengaud}, E. and {Asorey}, J. and {Avila}, S. and {Aviles}, A. and {Bailey}, S. and {Balaguera-Antol{\'\i}nez}, A. and {Ballester}, O. and {Baltay}, C. and {Bault}, A. and {Bautista}, J. and {Behera}, J. and {Beltran}, S.~F. and {BenZvi}, S. and {Beraldo e Silva}, L. and {Bermejo-Climent}, J.~R. and {Berti}, A. and {Besuner}, R. and {Beutler}, F. and {Bianchi}, D. and {Blake}, C. and {Blum}, R. and {Bolton}, A.~S. and {Brieden}, S. and {Brodzeller}, A. and {Brooks}, D. and {Brown}, Z. and {Buckley-Geer}, E. and {Burtin}, E. and {Cabayol-Garcia}, L. and {Cai}, Z. and {Canning}, R. and {Cardiel-Sas}, L. and {Carnero Rosell}, A. and {Castander}, F.~J. and {Cervantes-Cota}, J.~L. and {Chabanier}, S. and {Chaussidon}, E. and {Chaves-Montero}, J. and {Chen}, S. and {Chen}, X. and {Chuang}, C. and {Claybaugh}, T. and {Cole}, S. and {Cooper}, A.~P. and {Cuceu}, A. and {Davis}, T.~M. and {Dawson}, K. and {de Belsunce}, R. and {de la Cruz}, R. and {de la Macorra}, A. and {de Mattia}, A. and {Demina}, R. and {Demirbozan}, U. and {DeRose}, J. and {Dey}, A. and {Dey}, B. and {Dhungana}, G. and {Ding}, J. and {Ding}, Z. and {Doel}, P. and {Doshi}, R. and {Douglass}, K. and {Edge}, A. and {Eftekharzadeh}, S. and {Eisenstein}, D.~J. and {Elliott}, A. and {Escoffier}, S. and {Fagrelius}, P. and {Fan}, X. and {Fanning}, K. and {Fawcett}, V.~A. and {Ferraro}, S. and {Ereza}, J. and {Flaugher}, B. and {Font-Ribera}, A. and {Forero-S{\'a}nchez}, D. and {Forero-Romero}, J.~E. and {Frenk}, C.~S. and {G{\"a}nsicke}, B.~T. and {Garc{\'\i}a}, L. {\'A}. and {Garc{\'\i}a-Bellido}, J. and {Garcia-Quintero}, C. and {Garrison}, L.~H. and {Gil-Mar{\'\i}n}, H. and {Golden-Marx}, J. and {Gontcho A Gontcho}, S. and {Gonzalez-Morales}, A.~X. and {Gonzalez-Perez}, V. and {Gordon}, C. and {Graur}, O. and {Green}, D. and {Gruen}, D. and {Guy}, J. and {Hadzhiyska}, B. and {Hahn}, C. and {Han}, J.~J. and {Hanif}, M.~M.~S. and {Herrera-Alcantar}, H.~K. and {Honscheid}, K. and {Hou}, J. and {Howlett}, C. and {Huterer}, D. and {Ir{\v{s}}i{\v{c}}}, V. and {Ishak}, M. and {Jana}, A. and {Jiang}, L. and {Jimenez}, J. and {Jing}, Y.~P. and {Joudaki}, S. and {Jullo}, E. and {Joyce}, R. and {Juneau}, S. and {Kizhuprakkat}, N. and {Kara{\c{c}}ayl{\i}}, N.~G. and {Karim}, T. and {Kehoe}, R. and {Kent}, S. and {Khederlarian}, A. and {Kim}, S. and {Kirkby}, D. and {Kisner}, T. and {Kitaura}, F. and {Kneib}, J. and {Koposov}, S.~E. and {Kov{\'a}cs}, A. and {Kremin}, A. and {Krolewski}, A. and {L'Huillier}, B. and {Lahav}, O. and {Lambert}, A. and {Lamman}, C. and {Lan}, T. -W. and {Landriau}, M. and {Lang}, D. and {Lange}, J.~U. and {Lasker}, J. and {Le Guillou}, L. and {Leauthaud}, A. and {Levi}, M.~E. and {Li}, T.~S. and {Linder}, E. and {Lyons}, A. and {Magneville}, C. and {Manera}, M. and {Manser}, C.~J. and {Margala}, D. and {Martini}, P. and {McDonald}, P. and {Medina}, G.~E. and {Medina-Varela}, L. and {Meisner}, A. and {Mena-Fern{\'a}ndez}, J. and {Meneses-Rizo}, J. and {Mezcua}, M. and {Miquel}, R. and {Montero-Camacho}, P. and {Moon}, J. and {Moore}, S. and {Moustakas}, J. and {Mueller}, E. and {Mundet}, J. and {Mu{\~n}oz-Guti{\'e}rrez}, A. and {Myers}, A.~D. and {Nadathur}, S. and {Napolitano}, L. and {Neveux}, R. and {Newman}, J.~A. and {Nie}, J. and {Niz}, G. and {Norberg}, P. and {Noriega}, H.~E. and {Paillas}, E. and {Palanque-Delabrouille}, N. and {Palmese}, A. and {Zhiwei}, P. and {Parkinson}, D. and {Penmetsa}, S. and {Percival}, W.~J. and {P{\'e}rez-Fern{\'a}ndez}, A. and {P{\'e}rez-R{\`a}fols}, I. and {Pieri}, M. and {Poppett}, C. and {Porredon}, A. and {Prada}, F. and {Pucha}, R. and {Raichoor}, A. and {Ram{\'\i}rez-P{\'e}rez}, C.},
        title = "{Validation of the Scientific Program for the Dark Energy Spectroscopic Instrument}",
      journal = {\aj},
         year = 2024,
        month = feb,
       volume = {167},
       number = {2},
          eid = {62},
        pages = {62},
          doi = {10.3847/1538-3881/ad0b08},
archivePrefix = {arXiv},
       eprint = {2306.06307},
 primaryClass = {astro-ph.CO},
       adsurl = {https://ui.adsabs.harvard.edu/abs/2024AJ....167...62D}
}

@ARTICLE{DESI_DR1_BAO,
       author = {{DESI Collaboration} and {Adame}, A.~G. and {Aguilar}, J. and {Ahlen}, S. and {Alam}, S. and {Alexander}, D.~M. and {Alvarez}, M. and {Alves}, O. and {Anand}, A. and {Andrade}, U. and {Armengaud}, E. and {Avila}, S. and {Aviles}, A. and {Awan}, H. and {Bahr-Kalus}, B. and {Bailey}, S. and {Baltay}, C. and {Bault}, A. and {Behera}, J. and {BenZvi}, S. and {Bera}, A. and {Beutler}, F. and {Bianchi}, D. and {Blake}, C. and {Blum}, R. and {Brieden}, S. and {Brodzeller}, A. and {Brooks}, D. and {Buckley-Geer}, E. and {Burtin}, E. and {Calderon}, R. and {Canning}, R. and {Carnero Rosell}, A. and {Cereskaite}, R. and {Cervantes-Cota}, J.~L. and {Chabanier}, S. and {Chaussidon}, E. and {Chaves-Montero}, J. and {Chen}, S. and {Chen}, X. and {Claybaugh}, T. and {Cole}, S. and {Cuceu}, A. and {Davis}, T.~M. and {Dawson}, K. and {de la Macorra}, A. and {de Mattia}, A. and {Deiosso}, N. and {Dey}, A. and {Dey}, B. and {Ding}, Z. and {Doel}, P. and {Edelstein}, J. and {Eftekharzadeh}, S. and {Eisenstein}, D.~J. and {Elliott}, A. and {Fagrelius}, P. and {Fanning}, K. and {Ferraro}, S. and {Ereza}, J. and {Findlay}, N. and {Flaugher}, B. and {Font-Ribera}, A. and {Forero-S{\'a}nchez}, D. and {Forero-Romero}, J.~E. and {Frenk}, C.~S. and {Garcia-Quintero}, C. and {Gazta{\~n}aga}, E. and {Gil-Mar{\'\i}n}, H. and {Gontcho a Gontcho}, S. and {Gonzalez-Morales}, A.~X. and {Gonzalez-Perez}, V. and {Gordon}, C. and {Green}, D. and {Gruen}, D. and {Gsponer}, R. and {Gutierrez}, G. and {Guy}, J. and {Hadzhiyska}, B. and {Hahn}, C. and {Hanif}, M.~M.~S. and {Herrera-Alcantar}, H.~K. and {Honscheid}, K. and {Howlett}, C. and {Huterer}, D. and {Ir{\v{s}}i{\v{c}}}, V. and {Ishak}, M. and {Juneau}, S. and {Kara{\c{c}}ayl{\i}}, N.~G. and {Kehoe}, R. and {Kent}, S. and {Kirkby}, D. and {Kremin}, A. and {Krolewski}, A. and {Lai}, Y. and {Lan}, T. -W. and {Landriau}, M. and {Lang}, D. and {Lasker}, J. and {Le Goff}, J.~M. and {Le Guillou}, L. and {Leauthaud}, A. and {Levi}, M.~E. and {Li}, T.~S. and {Linder}, E. and {Lodha}, K. and {Magneville}, C. and {Manera}, M. and {Margala}, D. and {Martini}, P. and {Maus}, M. and {McDonald}, P. and {Medina-Varela}, L. and {Meisner}, A. and {Mena-Fern{\'a}ndez}, J. and {Miquel}, R. and {Moon}, J. and {Moore}, S. and {Moustakas}, J. and {Mueller}, E. and {Mu{\~n}oz-Guti{\'e}rrez}, A. and {Myers}, A.~D. and {Nadathur}, S. and {Napolitano}, L. and {Neveux}, R. and {Newman}, J.~A. and {Nguyen}, N.~M. and {Nie}, J. and {Niz}, G. and {Noriega}, H.~E. and {Padmanabhan}, N. and {Paillas}, E. and {Palanque-Delabrouille}, N. and {Pan}, J. and {Penmetsa}, S. and {Percival}, W.~J. and {Pieri}, M.~M. and {Pinon}, M. and {Poppett}, C. and {Porredon}, A. and {Prada}, F. and {P{\'e}rez-Fern{\'a}ndez}, A. and {P{\'e}rez-R{\`a}fols}, I. and {Rabinowitz}, D. and {Raichoor}, A. and {Ram{\'\i}rez-P{\'e}rez}, C. and {Ramirez-Solano}, S. and {Rashkovetskyi}, M. and {Ravoux}, C. and {Rezaie}, M. and {Rich}, J. and {Rocher}, A. and {Rockosi}, C. and {Roe}, N.~A. and {Rosado-Marin}, A. and {Ross}, A.~J. and {Rossi}, G. and {Ruggeri}, R. and {Ruhlmann-Kleider}, V. and {Samushia}, L. and {Sanchez}, E. and {Saulder}, C. and {Schlafly}, E.~F. and {Schlegel}, D. and {Schubnell}, M. and {Seo}, H. and {Shafieloo}, A. and {Sharples}, R. and {Silber}, J. and {Slosar}, A. and {Smith}, A. and {Sprayberry}, D. and {Tan}, T. and {Tarl{\'e}}, G. and {Taylor}, P. and {Trusov}, S. and {Ure{\~n}a-L{\'o}pez}, L.~A. and {Vaisakh}, R. and {Valcin}, D. and {Valdes}, F. and {Vargas-Maga{\~n}a}, M. and {Verde}, L. and {Walther}, M. and {Wang}, B. and {Wang}, M.~S. and {Weaver}, B.~A. and {Weaverdyck}, N. and {Wechsler}, R.~H. and {Weinberg}, D.~H. and {White}, M. and {Yu}, J. and {Yu}, Y. and {Yuan}, S. and {Y{\`e}che}, C. and {Zaborowski}, E.~A. and {Zarrouk}, P. and {Zhang}, H. and {Zhao}, C. and {Zhao}, R. and {Zhou}, R. and {Zhuang}, T.},
        title = "{DESI 2024 VI: cosmological constraints from the measurements of baryon acoustic oscillations}",
      journal = {\jcap},
         year = 2025,
        month = feb,
       volume = {2025},
       number = {2},
          eid = {021},
        pages = {021},
          doi = {10.1088/1475-7516/2025/02/021},
archivePrefix = {arXiv},
       eprint = {2404.03002},
 primaryClass = {astro-ph.CO},
       adsurl = {https://ui.adsabs.harvard.edu/abs/2025JCAP...02..021A}
}

@ARTICLE{DESI_DR1_fs,
       author = {{DESI Collaboration} and {Adame}, A.~G. and {Aguilar}, J. and {Ahlen}, S. and {Alam}, S. and {Alexander}, D.~M. and {Allende Prieto}, C. and {Alvarez}, M. and {Alves}, O. and {Anand}, A. and {Andrade}, U. and {Armengaud}, E. and {Avila}, S. and {Aviles}, A. and {Awan}, H. and {Bahr-Kalus}, B. and {Bailey}, S. and {Baltay}, C. and {Bault}, A. and {Behera}, J. and {BenZvi}, S. and {Beutler}, F. and {Bianchi}, D. and {Blake}, C. and {Blum}, R. and {Bonici}, M. and {Brieden}, S. and {Brodzeller}, A. and {Brooks}, D. and {Buckley-Geer}, E. and {Burtin}, E. and {Calderon}, R. and {Canning}, R. and {Carnero Rosell}, A. and {Cereskaite}, R. and {Cervantes-Cota}, J.~L. and {Chabanier}, S. and {Chaussidon}, E. and {Chaves-Montero}, J. and {Chebat}, D. and {Chen}, S. and {Chen}, X. and {Claybaugh}, T. and {Cole}, S. and {Cuceu}, A. and {Davis}, T.~M. and {Dawson}, K. and {de la Macorra}, A. and {de Mattia}, A. and {Deiosso}, N. and {Dey}, A. and {Dey}, B. and {Ding}, Z. and {Doel}, P. and {Edelstein}, J. and {Eftekharzadeh}, S. and {Eisenstein}, D.~J. and {Elbers}, W. and {Elliott}, A. and {Fagrelius}, P. and {Fanning}, K. and {Ferraro}, S. and {Ereza}, J. and {Findlay}, N. and {Flaugher}, B. and {Font-Ribera}, A. and {Forero-S{\'a}nchez}, D. and {Forero-Romero}, J.~E. and {Frenk}, C.~S. and {Garcia-Quintero}, C. and {Garrison}, L.~H. and {Gazta{\~n}aga}, E. and {Gil-Mar{\'\i}n}, H. and {Gontcho}, S. Gontcho A. and {Gonzalez-Morales}, A.~X. and {Gonzalez-Perez}, V. and {Gordon}, C. and {Green}, D. and {Gruen}, D. and {Gsponer}, R. and {Gutierrez}, G. and {Guy}, J. and {Hadzhiyska}, B. and {Hahn}, C. and {Hanif}, M.~M.~S. and {Herrera-Alcantar}, H.~K. and {Honscheid}, K. and {Howlett}, C. and {Huterer}, D. and {Ir{\v{s}}i{\v{c}}}, V. and {Ishak}, M. and {Joyce}, R. and {Juneau}, S. and {Kara{\c{c}}ayl{\i}}, N.~G. and {Kehoe}, R. and {Kent}, S. and {Kirkby}, D. and {Kong}, H. and {Koposov}, S.~E. and {Kremin}, A. and {Krolewski}, A. and {Lahav}, O. and {Lai}, Y. and {Lan}, T. -W. and {Landriau}, M. and {Lang}, D. and {Lasker}, J. and {Le Goff}, J.~M. and {Le Guillou}, L. and {Leauthaud}, A. and {Levi}, M.~E. and {Li}, T.~S. and {Lodha}, K. and {Magneville}, C. and {Manera}, M. and {Margala}, D. and {Martini}, P. and {Matthewson}, W. and {Maus}, M. and {McDonald}, P. and {Medina-Varela}, L. and {Meisner}, A. and {Mena-Fern{\'a}ndez}, J. and {Miquel}, R. and {Moon}, J. and {Moore}, S. and {Moustakas}, J. and {Mudur}, N. and {Mueller}, E. and {Mu{\~n}oz-Guti{\'e}rrez}, A. and {Myers}, A.~D. and {Nadathur}, S. and {Napolitano}, L. and {Neveux}, R. and {Newman}, J.~A. and {Nguyen}, N.~M. and {Nie}, J. and {Niz}, G. and {Noriega}, H.~E. and {Padmanabhan}, N. and {Paillas}, E. and {Palanque-Delabrouille}, N. and {Pan}, J. and {Penmetsa}, S. and {Percival}, W.~J. and {Pieri}, M.~M. and {Pinon}, M. and {Poppett}, C. and {Porredon}, A. and {Prada}, F. and {P{\'e}rez-Fern{\'a}ndez}, A. and {P{\'e}rez-R{\`a}fols}, I. and {Rabinowitz}, D. and {Raichoor}, A. and {Ram{\'\i}rez-P{\'e}rez}, C. and {Ramirez-Solano}, S. and {Rashkovetskyi}, M. and {Ravoux}, C. and {Rezaie}, M. and {Rich}, J. and {Rocher}, A. and {Rockosi}, C. and {Roe}, N.~A. and {Rosado-Marin}, A. and {Ross}, A.~J. and {Rossi}, G. and {Ruggeri}, R. and {Ruhlmann-Kleider}, V. and {Samushia}, L. and {Sanchez}, E. and {Saulder}, C. and {Schlafly}, E.~F. and {Schlegel}, D. and {Schubnell}, M. and {Seo}, H. and {Shafieloo}, A. and {Sharples}, R. and {Silber}, J. and {Slosar}, A. and {Smith}, A. and {Sprayberry}, D. and {Tan}, T. and {Tarl{\'e}}, G. and {Taylor}, P. and {Trusov}, S. and {Vaisakh}, R. and {Valcin}, D. and {Valdes}, F. and {Valogiannis}, G. and {Vargas-Maga{\~n}a}, M. and {Verde}, L. and {Walther}, M. and {Wang}, B. and {Wang}, M.~S. and {Weaver}, B.~A. and {Weaverdyck}, N. and {Wechsler}, R.~H. and {Weinberg}, D.~H. and {White}, M. and {Wilson}, M.~J. and {Yi}, L.},
        title = "{DESI 2024 VII: cosmological constraints from the full-shape modeling of clustering measurements}",
      journal = {\jcap},
         year = 2025,
        month = jul,
       volume = {2025},
       number = {7},
          eid = {028},
        pages = {028},
          doi = {10.1088/1475-7516/2025/07/028},
archivePrefix = {arXiv},
       eprint = {2411.12022},
 primaryClass = {astro-ph.CO},
       adsurl = {https://ui.adsabs.harvard.edu/abs/2025JCAP...07..028A}
}

@ARTICLE{DESI_DR2_BAO,
      title="{DESI DR2 Results II: Measurements of Baryon Acoustic Oscillations and Cosmological Constraints}", 
      author={{DESI Collaboration} and M. Abdul-Karim and J. Aguilar and S. Ahlen and S. Alam and L. Allen and C. Allende Prieto and O. Alves and A. Anand and U. Andrade and E. Armengaud and A. Aviles and S. Bailey and C. Baltay and P. Bansal and A. Bault and J. Behera and S. BenZvi and D. Bianchi and C. Blake and S. Brieden and A. Brodzeller and D. Brooks and E. Buckley-Geer and E. Burtin and R. Calderon and R. Canning and A. Carnero Rosell and P. Carrilho and L. Casas and F. J. Castander and R. Cereskaite and M. Charles and E. Chaussidon and J. Chaves-Montero and D. Chebat and X. Chen and T. Claybaugh and S. Cole and A. P. Cooper and A. Cuceu and K. S. Dawson and A. de la Macorra and A. de Mattia and N. Deiosso and J. Della Costa and R. Demina and A. Dey and B. Dey and Z. Ding and P. Doel and J. Edelstein and D. J. Eisenstein and W. Elbers and P. Fagrelius and K. Fanning and E. Fernández-García and S. Ferraro and A. Font-Ribera and J. E. Forero-Romero and C. S. Frenk and C. Garcia-Quintero and L. H. Garrison and E. Gaztañaga and H. Gil-Marín and S. Gontcho A Gontcho and D. Gonzalez and A. X. Gonzalez-Morales and C. Gordon and D. Green and G. Gutierrez and J. Guy and B. Hadzhiyska and C. Hahn and S. He and M. Herbold and H. K. Herrera-Alcantar and M. Ho and K. Honscheid and C. Howlett and D. Huterer and M. Ishak and S. Juneau and N. V. Kamble and N. G. Karaçaylı and R. Kehoe and S. Kent and A. G. Kim and D. Kirkby and T. Kisner and S. E. Koposov and A. Kremin and A. Krolewski and O. Lahav and C. Lamman and M. Landriau and D. Lang and J. Lasker and J. M. Le Goff and L. Le Guillou and A. Leauthaud and M. E. Levi and Q. Li and T. S. Li and K. Lodha and M. Lokken and F. Lozano-Rodríguez and C. Magneville and M. Manera and P. Martini and W. L. Matthewson and A. Meisner and J. Mena-Fernández and A. Menegas and T. Mergulhão and R. Miquel and J. Moustakas and A. Muñoz-Gutiérrez and D. Muñoz-Santos and A. D. Myers and S. Nadathur and K. Naidoo and L. Napolitano and J. A. Newman and G. Niz and H. E. Noriega and E. Paillas and N. Palanque-Delabrouille and J. Pan and J. Peacock and Marcos Pellejero Ibanez and W. J. Percival and A. Pérez-Fernández and I. Pérez-Ràfols and M. M. Pieri and C. Poppett and F. Prada and D. Rabinowitz and A. Raichoor and C. Ramírez-Pérez and M. Rashkovetskyi and C. Ravoux and J. Rich and A. Rocher and C. Rockosi and J. Rohlf and J. O. Román-Herrera and A. J. Ross and G. Rossi and R. Ruggeri and V. Ruhlmann-Kleider and L. Samushia and E. Sanchez and N. Sanders and D. Schlegel and M. Schubnell and H. Seo and A. Shafieloo and R. Sharples and J. Silber and F. Sinigaglia and D. Sprayberry and T. Tan and G. Tarlé and P. Taylor and W. Turner and L. A. Ureña-López and R. Vaisakh and F. Valdes and G. Valogiannis and M. Vargas-Magaña and L. Verde and M. Walther and B. A. Weaver and D. H. Weinberg and M. White and M. Wolfson and C. Yèche and J. Yu and E. A. Zaborowski and P. Zarrouk and Z. Zhai and H. Zhang and C. Zhao and G. B. Zhao and R. Zhou and H. Zou},
      year=2025,
      journal={arXiv:2503.14738},
      doi={10.48550/arXiv.2503.14738}, 
}

@ARTICLE{Euclid_Q1_overview,
       author = {{Euclid Collaboration} and {Aussel}, H. and {Tereno}, I. and {Schirmer}, M. and {Alguero}, G. and {Altieri}, B. and {Balbinot}, E. and {de Boer}, T. and {Casenove}, P. and {Corcho-Caballero}, P. and {Furusawa}, H. and {Furusawa}, J. and {Hudson}, M.~J. and {Jahnke}, K. and {Libet}, G. and {Macias-Perez}, J. and {Masoumzadeh}, N. and {Mohr}, J.~J. and {Odier}, J. and {Scott}, D. and {Vassallo}, T. and {Verdoes Kleijn}, G. and {Zacchei}, A. and {Aghanim}, N. and {Amara}, A. and {Andreon}, S. and {Auricchio}, N. and {Awan}, S. and {Azzollini}, R. and {Baccigalupi}, C. and {Baldi}, M. and {Balestra}, A. and {Bardelli}, S. and {Basset}, A. and {Battaglia}, P. and {Belikov}, A.~N. and {Bender}, R. and {Biviano}, A. and {Bonchi}, A. and {Bonino}, D. and {Branchini}, E. and {Brescia}, M. and {Brinchmann}, J. and {Camera}, S. and {Ca{\~n}as-Herrera}, G. and {Capobianco}, V. and {Carbone}, C. and {Cardone}, V.~F. and {Carretero}, J. and {Casas}, S. and {Castander}, F.~J. and {Castellano}, M. and {Castignani}, G. and {Cavuoti}, S. and {Chambers}, K.~C. and {Cimatti}, A. and {Colodro-Conde}, C. and {Congedo}, G. and {Conselice}, C.~J. and {Conversi}, L. and {Copin}, Y. and {Courbin}, F. and {Courtois}, H.~M. and {Cropper}, M. and {Cuby}, J. -G. and {Da Silva}, A. and {da Silva}, R. and {Degaudenzi}, H. and {de Jong}, J.~T.~A. and {De Lucia}, G. and {Di Giorgio}, A.~M. and {Dinis}, J. and {Dolding}, C. and {Dole}, H. and {Douspis}, M. and {Dubath}, F. and {Duncan}, C.~A.~J. and {Dupac}, X. and {Dusini}, S. and {Ealet}, A. and {Escoffier}, S. and {Fabricius}, M. and {Farina}, M. and {Farinelli}, R. and {Faustini}, F. and {Ferriol}, S. and {Fotopoulou}, S. and {Fourmanoit}, N. and {Frailis}, M. and {Franceschi}, E. and {Franzetti}, P. and {Galeotta}, S. and {George}, K. and {Gillard}, W. and {Gillis}, B. and {Giocoli}, C. and {G{\'o}mez-Alvarez}, P. and {Gracia-Carpio}, J. and {Granett}, B.~R. and {Grazian}, A. and {Grupp}, F. and {Guzzo}, L. and {Gwyn}, S. and {Haugan}, S.~V.~H. and {Herent}, O. and {Hoar}, J. and {Hoekstra}, H. and {Holliman}, M.~S. and {Holmes}, W. and {Hook}, I.~M. and {Hormuth}, F. and {Hornstrup}, A. and {Hudelot}, P. and {Ili{\'c}}, S. and {Jhabvala}, M. and {Joachimi}, B. and {Keih{\"a}nen}, E. and {Kermiche}, S. and {Kiessling}, A. and {Kubik}, B. and {Kuijken}, K. and {K{\"u}mmel}, M. and {Kunz}, M. and {Kurki-Suonio}, H. and {Lahav}, O. and {Le Boulc'h}, Q. and {Le Brun}, A.~M.~C. and {Le Mignant}, D. and {Liebing}, P. and {Ligori}, S. and {Lilje}, P.~B. and {Lindholm}, V. and {Lloro}, I. and {Mainetti}, G. and {Maino}, D. and {Maiorano}, E. and {Mansutti}, O. and {Marcin}, S. and {Marggraf}, O. and {Markovic}, K. and {Martinelli}, M. and {Martinet}, N. and {Marulli}, F. and {Massey}, R. and {Maurogordato}, S. and {McCracken}, H.~J. and {Medinaceli}, E. and {Mei}, S. and {Melchior}, M. and {Mellier}, Y. and {Meneghetti}, M. and {Merlin}, E. and {Meylan}, G. and {Mora}, A. and {Moresco}, M. and {Morris}, P.~W. and {Moscardini}, L. and {Mourre}, S. and {Nakajima}, R. and {Neissner}, C. and {Nichol}, R.~C. and {Niemi}, S. -M. and {Nightingale}, J.~W. and {Nutma}, T. and {Padilla}, C. and {Paltani}, S. and {Pasian}, F. and {Peacock}, J.~A. and {Pedersen}, K. and {Percival}, W.~J. and {Pettorino}, V. and {Pires}, S. and {Polenta}, G. and {Pollack}, J.~E. and {Poncet}, M. and {Popa}, L.~A. and {Pozzetti}, L. and {Racca}, G.~D. and {Raison}, F. and {Rebolo}, R. and {Renzi}, A. and {Rhodes}, J. and {Riccio}, G. and {Rix}, H. -W. and {Romelli}, E. and {Roncarelli}, M. and {Rossetti}, E. and {Rusholme}, B. and {Saglia}, R. and {Sakr}, Z. and {S{\'a}nchez}, A.~G. and {Sapone}, D. and {Sartoris}, B. and {Sauvage}, M. and {Schewtschenko}, J.~A. and {Schneider}, P. and {Scodeggio}, M. and {Secroun}, A. and {Sefusatti}, E. and {Seidel}, G.},
        title = "{Euclid Quick Data Release (Q1) -- Data release overview}",
      journal = {arXiv:2503.15302},
         year = 2025,
        month = mar,
          doi = {10.48550/arXiv.2503.15302}
}

@ARTICLE{Euclid_EWS,
       author = {{Euclid Collaboration} and {Scaramella}, R. and {Amiaux}, J. and {Mellier}, Y. and {Burigana}, C. and {Carvalho}, C.~S. and {Cuillandre}, J. -C. and {Da Silva}, A. and {Derosa}, A. and {Dinis}, J. and {Maiorano}, E. and {Maris}, M. and {Tereno}, I. and {Laureijs}, R. and {Boenke}, T. and {Buenadicha}, G. and {Dupac}, X. and {Gaspar Venancio}, L.~M. and {G{\'o}mez-{\'A}lvarez}, P. and {Hoar}, J. and {Lorenzo Alvarez}, J. and {Racca}, G.~D. and {Saavedra-Criado}, G. and {Schwartz}, J. and {Vavrek}, R. and {Schirmer}, M. and {Aussel}, H. and {Azzollini}, R. and {Cardone}, V.~F. and {Cropper}, M. and {Ealet}, A. and {Garilli}, B. and {Gillard}, W. and {Granett}, B.~R. and {Guzzo}, L. and {Hoekstra}, H. and {Jahnke}, K. and {Kitching}, T. and {Maciaszek}, T. and {Meneghetti}, M. and {Miller}, L. and {Nakajima}, R. and {Niemi}, S.~M. and {Pasian}, F. and {Percival}, W.~J. and {Pottinger}, S. and {Sauvage}, M. and {Scodeggio}, M. and {Wachter}, S. and {Zacchei}, A. and {Aghanim}, N. and {Amara}, A. and {Auphan}, T. and {Auricchio}, N. and {Awan}, S. and {Balestra}, A. and {Bender}, R. and {Bodendorf}, C. and {Bonino}, D. and {Branchini}, E. and {Brau-Nogue}, S. and {Brescia}, M. and {Candini}, G.~P. and {Capobianco}, V. and {Carbone}, C. and {Carlberg}, R.~G. and {Carretero}, J. and {Casas}, R. and {Castander}, F.~J. and {Castellano}, M. and {Cavuoti}, S. and {Cimatti}, A. and {Cledassou}, R. and {Congedo}, G. and {Conselice}, C.~J. and {Conversi}, L. and {Copin}, Y. and {Corcione}, L. and {Costille}, A. and {Courbin}, F. and {Degaudenzi}, H. and {Douspis}, M. and {Dubath}, F. and {Duncan}, C.~A.~J. and {Dusini}, S. and {Farrens}, S. and {Ferriol}, S. and {Fosalba}, P. and {Fourmanoit}, N. and {Frailis}, M. and {Franceschi}, E. and {Franzetti}, P. and {Fumana}, M. and {Gillis}, B. and {Giocoli}, C. and {Grazian}, A. and {Grupp}, F. and {Haugan}, S.~V.~H. and {Holmes}, W. and {Hormuth}, F. and {Hudelot}, P. and {Kermiche}, S. and {Kiessling}, A. and {Kilbinger}, M. and {Kohley}, R. and {Kubik}, B. and {K{\"u}mmel}, M. and {Kunz}, M. and {Kurki-Suonio}, H. and {Lahav}, O. and {Ligori}, S. and {Lilje}, P.~B. and {Lloro}, I. and {Mansutti}, O. and {Marggraf}, O. and {Markovic}, K. and {Marulli}, F. and {Massey}, R. and {Maurogordato}, S. and {Melchior}, M. and {Merlin}, E. and {Meylan}, G. and {Mohr}, J.~J. and {Moresco}, M. and {Morin}, B. and {Moscardini}, L. and {Munari}, E. and {Nichol}, R.~C. and {Padilla}, C. and {Paltani}, S. and {Peacock}, J. and {Pedersen}, K. and {Pettorino}, V. and {Pires}, S. and {Poncet}, M. and {Popa}, L. and {Pozzetti}, L. and {Raison}, F. and {Rebolo}, R. and {Rhodes}, J. and {Rix}, H. -W. and {Roncarelli}, M. and {Rossetti}, E. and {Saglia}, R. and {Schneider}, P. and {Schrabback}, T. and {Secroun}, A. and {Seidel}, G. and {Serrano}, S. and {Sirignano}, C. and {Sirri}, G. and {Skottfelt}, J. and {Stanco}, L. and {Starck}, J.~L. and {Tallada-Cresp{\'\i}}, P. and {Tavagnacco}, D. and {Taylor}, A.~N. and {Teplitz}, H.~I. and {Toledo-Moreo}, R. and {Torradeflot}, F. and {Trifoglio}, M. and {Valentijn}, E.~A. and {Valenziano}, L. and {Verdoes Kleijn}, G.~A. and {Wang}, Y. and {Welikala}, N. and {Weller}, J. and {Wetzstein}, M. and {Zamorani}, G. and {Zoubian}, J. and {Andreon}, S. and {Baldi}, M. and {Bardelli}, S. and {Boucaud}, A. and {Camera}, S. and {Di Ferdinando}, D. and {Fabbian}, G. and {Farinelli}, R. and {Galeotta}, S. and {Graci{\'a}-Carpio}, J. and {Maino}, D. and {Medinaceli}, E. and {Mei}, S. and {Neissner}, C. and {Polenta}, G. and {Renzi}, A. and {Romelli}, E. and {Rosset}, C. and {Sureau}, F. and {Tenti}, M. and {Vassallo}, T. and {Zucca}, E. and {Baccigalupi}, C. and {Balaguera-Antol{\'\i}nez}, A. and {Battaglia}, P. and {Biviano}, A. and {Borgani}, S. and {Bozzo}, E. and {Cabanac}, R. and {Cappi}, A.},
        title = "{Euclid preparation. I. The Euclid Wide Survey}",
      journal = {\aap},
         year = 2022,
        month = jun,
       volume = {662},
          eid = {A112},
        pages = {A112},
          doi = {10.1051/0004-6361/202141938},
archivePrefix = {arXiv},
       eprint = {2108.01201},
 primaryClass = {astro-ph.CO},
       adsurl = {https://ui.adsabs.harvard.edu/abs/2022A&A...662A.112E}
}

@ARTICLE{White_1978,
       author = {{White}, S.~D.~M. and {Rees}, M.~J.},
        title = "{Core condensation in heavy halos: a two-stage theory for galaxy formation and clustering.}",
      journal = {\mnras},
         year = 1978,
        month = may,
       volume = {183},
        pages = {341-358},
          doi = {10.1093/mnras/183.3.341},
       adsurl = {https://ui.adsabs.harvard.edu/abs/1978MNRAS.183..341W}
}

@ARTICLE{Blumenthal_1984,
       author = {{Blumenthal}, G.~R. and {Faber}, S.~M. and {Primack}, J.~R. and {Rees}, M.~J.},
        title = "{Formation of galaxies and large-scale structure with cold dark matter.}",
      journal = {\nat},
         year = 1984,
        month = oct,
       volume = {311},
        pages = {517-525},
          doi = {10.1038/311517a0},
       adsurl = {https://ui.adsabs.harvard.edu/abs/1984Natur.311..517B}
}

@ARTICLE{excursion_set,
       author = {{Bond}, J.~R. and {Cole}, S. and {Efstathiou}, G. and {Kaiser}, N.},
        title = "{Excursion Set Mass Functions for Hierarchical Gaussian Fluctuations}",
      journal = {\apj},
         year = 1991,
        month = oct,
       volume = {379},
        pages = {440},
          doi = {10.1086/170520},
       adsurl = {https://ui.adsabs.harvard.edu/abs/1991ApJ...379..440B}
}

@ARTICLE{press_schechter,
       author = {{Press}, William H. and {Schechter}, Paul},
        title = "{Formation of Galaxies and Clusters of Galaxies by Self-Similar Gravitational Condensation}",
      journal = {\apj},
         year = 1974,
        month = feb,
       volume = {187},
        pages = {425-438},
          doi = {10.1086/152650},
       adsurl = {https://ui.adsabs.harvard.edu/abs/1974ApJ...187..425P}
}

@ARTICLE{Lacey_Cole_1993,
       author = {{Lacey}, Cedric and {Cole}, Shaun},
        title = "{Merger rates in hierarchical models of galaxy formation}",
      journal = {\mnras},
         year = 1993,
        month = jun,
       volume = {262},
       number = {3},
        pages = {627-649},
          doi = {10.1093/mnras/262.3.627},
       adsurl = {https://ui.adsabs.harvard.edu/abs/1993MNRAS.262..627L}
}

@ARTICLE{Lemson_Kauffmann_1999,
       author = {{Lemson}, Gerard and {Kauffmann}, Guinevere},
        title = "{Environmental influences on dark matter haloes and consequences for the galaxies within them}",
      journal = {\mnras},
         year = 1999,
        month = jan,
       volume = {302},
       number = {1},
        pages = {111-117},
          doi = {10.1046/j.1365-8711.1999.02090.x},
       adsurl = {https://ui.adsabs.harvard.edu/abs/1999MNRAS.302..111L}
}

@article{Sheth_2004,
   title={On the environmental dependence of halo formation},
   volume={350},
   ISSN={1365-2966},
   url={http://dx.doi.org/10.1111/j.1365-2966.2004.07733.x},
   DOI={10.1111/j.1365-2966.2004.07733.x},
   number={4},
   journal={Monthly Notices of the Royal Astronomical Society},
   publisher={Oxford University Press (OUP)},
   author={Sheth, Ravi K. and Tormen, Giuseppe},
   year={2004},
   month=jun, pages={1385–1390} }

@ARTICLE{Gao_2005,
       author = {{Gao}, Liang and {Springel}, Volker and {White}, Simon D.~M.},
        title = "{The age dependence of halo clustering}",
      journal = {\mnras},
         year = 2005,
        month = oct,
       volume = {363},
       number = {1},
        pages = {L66-L70},
          doi = {10.1111/j.1745-3933.2005.00084.x},
archivePrefix = {arXiv},
       eprint = {astro-ph/0506510},
 primaryClass = {astro-ph},
       adsurl = {https://ui.adsabs.harvard.edu/abs/2005MNRAS.363L..66G}
}

@ARTICLE{Millennium,
       author = {{Springel}, Volker and {White}, Simon D.~M. and {Jenkins}, Adrian and {Frenk}, Carlos S. and {Yoshida}, Naoki and {Gao}, Liang and {Navarro}, Julio and {Thacker}, Robert and {Croton}, Darren and {Helly}, John and {Peacock}, John A. and {Cole}, Shaun and {Thomas}, Peter and {Couchman}, Hugh and {Evrard}, August and {Colberg}, J{\"o}rg and {Pearce}, Frazer},
        title = "{Simulations of the formation, evolution and clustering of galaxies and quasars}",
      journal = {\nat},
         year = 2005,
        month = jun,
       volume = {435},
       number = {7042},
        pages = {629-636},
          doi = {10.1038/nature03597},
archivePrefix = {arXiv},
       eprint = {astro-ph/0504097},
 primaryClass = {astro-ph},
       adsurl = {https://ui.adsabs.harvard.edu/abs/2005Natur.435..629S}
}

@ARTICLE{Gao_2007,
       author = {{Gao}, Liang and {White}, Simon D.~M.},
        title = "{Assembly bias in the clustering of dark matter haloes}",
      journal = {\mnras},
         year = 2007,
        month = apr,
       volume = {377},
       number = {1},
        pages = {L5-L9},
          doi = {10.1111/j.1745-3933.2007.00292.x},
archivePrefix = {arXiv},
       eprint = {astro-ph/0611921},
 primaryClass = {astro-ph},
       adsurl = {https://ui.adsabs.harvard.edu/abs/2007MNRAS.377L...5G}
}

@ARTICLE{Wechsler_2006,
       author = {{Wechsler}, Risa H. and {Zentner}, Andrew R. and {Bullock}, James S. and {Kravtsov}, Andrey V. and {Allgood}, Brandon},
        title = "{The Dependence of Halo Clustering on Halo Formation History, Concentration, and Occupation}",
      journal = {\apj},
         year = 2006,
        month = nov,
       volume = {652},
       number = {1},
        pages = {71-84},
          doi = {10.1086/507120},
archivePrefix = {arXiv},
       eprint = {astro-ph/0512416},
 primaryClass = {astro-ph},
       adsurl = {https://ui.adsabs.harvard.edu/abs/2006ApJ...652...71W}
}

@ARTICLE{Angulo_2008,
       author = {{Angulo}, R.~E. and {Baugh}, C.~M. and {Lacey}, C.~G.},
        title = "{The assembly bias of dark matter haloes to higher orders}",
      journal = {\mnras},
         year = 2008,
        month = jun,
       volume = {387},
       number = {2},
        pages = {921-932},
          doi = {10.1111/j.1365-2966.2008.13304.x},
archivePrefix = {arXiv},
       eprint = {0712.2280},
 primaryClass = {astro-ph},
       adsurl = {https://ui.adsabs.harvard.edu/abs/2008MNRAS.387..921A}
}

@article{Zhu_2006,
   title={The Dependence of the Occupation of Galaxies on the Halo Formation Time},
   volume={639},
   ISSN={1538-4357},
   url={http://dx.doi.org/10.1086/501501},
   DOI={10.1086/501501},
   number={1},
   journal={The Astrophysical Journal},
   publisher={American Astronomical Society},
   author={Zhu, Guangtun and Zheng, Zheng and Lin, W. P. and Jing, Y. P. and Kang, Xi and Gao, Liang},
   year={2006},
   month=feb, pages={L5–L8} }

@article{Zu_2008,
   title={Environmental Effects on Real‐Space and Redshift‐Space Galaxy Clustering},
   volume={686},
   ISSN={1538-4357},
   url={http://dx.doi.org/10.1086/591071},
   DOI={10.1086/591071},
   number={1},
   journal={The Astrophysical Journal},
   publisher={American Astronomical Society},
   author={Zu, Ying and Zheng, Zheng and Zhu, Guangtun and Jing, Y. P.},
   year={2008},
   month=oct, pages={41–52} }

@ARTICLE{Chaves_2016,
       author = {{Chaves-Montero}, Jon{\'a}s and {Angulo}, Raul E. and {Schaye}, Joop and {Schaller}, Matthieu and {Crain}, Robert A. and {Furlong}, Michelle and {Theuns}, Tom},
        title = "{Subhalo abundance matching and assembly bias in the EAGLE simulation}",
      journal = {\mnras},
         year = 2016,
        month = aug,
       volume = {460},
       number = {3},
        pages = {3100-3118},
          doi = {10.1093/mnras/stw1225},
archivePrefix = {arXiv},
       eprint = {1507.01948},
 primaryClass = {astro-ph.GA},
       adsurl = {https://ui.adsabs.harvard.edu/abs/2016MNRAS.460.3100C}
}

@article{Borzyszkowski_2017,
   title={ZOMG – I. How the cosmic web inhibits halo growth and generates assembly bias},
   volume={469},
   ISSN={1365-2966},
   url={http://dx.doi.org/10.1093/mnras/stx873},
   DOI={10.1093/mnras/stx873},
   number={1},
   journal={Monthly Notices of the Royal Astronomical Society},
   publisher={Oxford University Press (OUP)},
   author={Borzyszkowski, Mikolaj and Porciani, Cristiano and Romano-Díaz, Emilio and Garaldi, Enrico},
   year={2017},
   month=apr, pages={594–611} }

@ARTICLE{Tojeiro_2017,
       author = {{Tojeiro}, Rita and {Eardley}, Elizabeth and {Peacock}, John A. and {Norberg}, Peder and {Alpaslan}, Mehmet and {Driver}, Simon P. and {Henriques}, Bruno and {Hopkins}, Andrew M. and {Kafle}, Prajwal R. and {Robotham}, Aaron S.~G. and {Thomas}, Peter and {Tonini}, Chiara and {Wild}, Vivienne},
        title = "{Galaxy and Mass Assembly (GAMA): halo formation times and halo assembly bias on the cosmic web}",
      journal = {\mnras},
         year = 2017,
        month = sep,
       volume = {470},
       number = {3},
        pages = {3720-3741},
          doi = {10.1093/mnras/stx1466},
archivePrefix = {arXiv},
       eprint = {1612.08595},
 primaryClass = {astro-ph.CO},
       adsurl = {https://ui.adsabs.harvard.edu/abs/2017MNRAS.470.3720T}
}

@article{Yang_2017,
   title={Revealing the Cosmic Web-dependent Halo Bias},
   volume={848},
   ISSN={1538-4357},
   url={http://dx.doi.org/10.3847/1538-4357/aa8c7a},
   DOI={10.3847/1538-4357/aa8c7a},
   number={1},
   journal={The Astrophysical Journal},
   publisher={American Astronomical Society},
   author={Yang, Xiaohu and Zhang, Youcai and Lu, Tianhuan and Wang, Huiyuan and Shi, Feng and Tweed, Dylan and Li, Shijie and Luo, Wentao and Lu, Yi and Yang, Lei},
   year={2017},
   month=oct, pages={60} }

@article{Ramakrishnan_2019,
   title={Cosmic web anisotropy is the primary indicator of halo assembly bias},
   volume={489},
   ISSN={1365-2966},
   url={http://dx.doi.org/10.1093/mnras/stz2344},
   DOI={10.1093/mnras/stz2344},
   number={3},
   journal={Monthly Notices of the Royal Astronomical Society},
   publisher={Oxford University Press (OUP)},
   author={Ramakrishnan, Sujatha and Paranjape, Aseem and Hahn, Oliver and Sheth, Ravi K},
   year={2019},
   month=aug, pages={2977–2996} }

@article{Wu_2008,
   title={The Effects of Halo Assembly Bias on Self‐Calibration in Galaxy Cluster Surveys},
   volume={688},
   ISSN={1538-4357},
   url={http://dx.doi.org/10.1086/591929},
   DOI={10.1086/591929},
   number={2},
   journal={The Astrophysical Journal},
   publisher={American Astronomical Society},
   author={Wu, Hao‐Yi and Rozo, Eduardo and Wechsler, Risa H.},
   year={2008},
   month=dec, pages={729–741} }

@article{McCarthy_2019,
   title={The effects of galaxy assembly bias on the inference of growth rate from redshift-space distortions},
   volume={487},
   ISSN={1365-2966},
   url={http://dx.doi.org/10.1093/mnras/stz1461},
   DOI={10.1093/mnras/stz1461},
   number={2},
   journal={Monthly Notices of the Royal Astronomical Society},
   publisher={Oxford University Press (OUP)},
   author={McCarthy, Kevin S and Zheng, Zheng and Guo, Hong},
   year={2019},
   month=jun, pages={2424–2440} }

@article{Hadzhiyska_2021,
   title={Galaxy assembly bias and large-scale distribution: a comparison between IllustrisTNG and a semi-analytic model},
   volume={508},
   ISSN={1365-2966},
   url={http://dx.doi.org/10.1093/mnras/stab2564},
   DOI={10.1093/mnras/stab2564},
   number={1},
   journal={Monthly Notices of the Royal Astronomical Society},
   publisher={Oxford University Press (OUP)},
   author={Hadzhiyska, Boryana and Liu, Sonya and Somerville, Rachel S and Gabrielpillai, Austen and Bose, Sownak and Eisenstein, Daniel and Hernquist, Lars},
   year={2021},
   month=sep, pages={698–718} }

@ARTICLE{Montero_2024,
       author = {{Montero-Dorta}, Antonio D. and {Rodriguez}, Facundo},
        title = "{The dependence of assembly bias on the cosmic web}",
      journal = {\mnras},
         year = 2024,
        month = jun,
       volume = {531},
       number = {1},
        pages = {290-303},
          doi = {10.1093/mnras/stae796},
archivePrefix = {arXiv},
       eprint = {2309.12401},
 primaryClass = {astro-ph.GA},
       adsurl = {https://ui.adsabs.harvard.edu/abs/2024MNRAS.531..290M}
}

@article{Zehavi_2018,
   title={The Impact of Assembly Bias on the Galaxy Content of Dark Matter Halos},
   volume={853},
   ISSN={1538-4357},
   url={http://dx.doi.org/10.3847/1538-4357/aaa54a},
   DOI={10.3847/1538-4357/aaa54a},
   number={1},
   journal={The Astrophysical Journal},
   publisher={American Astronomical Society},
   author={Zehavi, Idit and Contreras, Sergio and Padilla, Nelson and Smith, Nicholas J. and Baugh, Carlton M. and Norberg, Peder},
   year={2018},
   month=jan, pages={84} }

@article{Artale_2018,
   title={The impact of assembly bias on the halo occupation in hydrodynamical simulations},
   volume={480},
   ISSN={1365-2966},
   url={http://dx.doi.org/10.1093/mnras/sty2110},
   DOI={10.1093/mnras/sty2110},
   number={3},
   journal={Monthly Notices of the Royal Astronomical Society},
   publisher={Oxford University Press (OUP)},
   author={Artale, M Celeste and Zehavi, Idit and Contreras, Sergio and Norberg, Peder},
   year={2018},
   month=aug, pages={3978–3992} }

@ARTICLE{Perez_2024,
       author = {{Perez}, Noelia R. and {Pereyra}, Luis A. and {Coldwell}, Georgina and {Rodriguez}, Facundo and {Alfaro}, Ignacio G. and {Ruiz}, Andr{\'e}s N.},
        title = "{Characterizing HOD in filaments and nodes of the cosmic web}",
      journal = {\mnras},
         year = 2024,
        month = feb,
       volume = {528},
       number = {2},
        pages = {3186-3197},
          doi = {10.1093/mnras/stae188},
archivePrefix = {arXiv},
       eprint = {2310.19928},
 primaryClass = {astro-ph.GA},
       adsurl = {https://ui.adsabs.harvard.edu/abs/2024MNRAS.528.3186P}
}

@ARTICLE{Zhang_2025,
       author = {{Zhang}, Ziwen and {Chen}, Yangyao and {Rong}, Yu and {Wang}, Huiyuan and {Mo}, Houjun and {Luo}, Xiong and {Li}, Hao},
        title = "{Unexpected clustering pattern in dwarf galaxies challenges formation models}",
      journal = {\nat},
         year = 2025,
        month = jun,
       volume = {642},
       number = {8066},
        pages = {47-52},
          doi = {10.1038/s41586-025-08965-5},
archivePrefix = {arXiv},
       eprint = {2504.03305},
 primaryClass = {astro-ph.CO},
       adsurl = {https://ui.adsabs.harvard.edu/abs/2025Natur.642...47Z}
}

@article{Vakili_2019,
   title={How Are Galaxies Assigned to Halos? Searching for Assembly Bias in the SDSS Galaxy Clustering},
   volume={872},
   ISSN={1538-4357},
   url={http://dx.doi.org/10.3847/1538-4357/aaf1a1},
   DOI={10.3847/1538-4357/aaf1a1},
   number={1},
   journal={The Astrophysical Journal},
   publisher={American Astronomical Society},
   author={Vakili, Mohammadjavad and Hahn, ChangHoon},
   year={2019},
   month=feb, pages={115} }

@ARTICLE{TNG_Method_1,
       author = {{Weinberger}, Rainer and {Springel}, Volker and {Hernquist}, Lars and {Pillepich}, Annalisa and {Marinacci}, Federico and {Pakmor}, R{\"u}diger and {Nelson}, Dylan and {Genel}, Shy and {Vogelsberger}, Mark and {Naiman}, Jill and {Torrey}, Paul},
        title = "{Simulating galaxy formation with black hole driven thermal and kinetic feedback}",
      journal = {\mnras},
         year = 2017,
        month = mar,
       volume = {465},
       number = {3},
        pages = {3291-3308},
          doi = {10.1093/mnras/stw2944},
archivePrefix = {arXiv},
       eprint = {1607.03486},
 primaryClass = {astro-ph.GA},
       adsurl = {https://ui.adsabs.harvard.edu/abs/2017MNRAS.465.3291W}
}

@ARTICLE{TNG_Method_2,
       author = {{Pillepich}, Annalisa and {Springel}, Volker and {Nelson}, Dylan and {Genel}, Shy and {Naiman}, Jill and {Pakmor}, R{\"u}diger and {Hernquist}, Lars and {Torrey}, Paul and {Vogelsberger}, Mark and {Weinberger}, Rainer and {Marinacci}, Federico},
        title = "{Simulating galaxy formation with the IllustrisTNG model}",
      journal = {\mnras},
         year = 2018,
        month = jan,
       volume = {473},
       number = {3},
        pages = {4077-4106},
          doi = {10.1093/mnras/stx2656},
archivePrefix = {arXiv},
       eprint = {1703.02970},
 primaryClass = {astro-ph.GA},
       adsurl = {https://ui.adsabs.harvard.edu/abs/2018MNRAS.473.4077P}
}

@article{Pillepich_2017,
   title={First results from the IllustrisTNG simulations: the stellar mass content of groups and clusters of galaxies},
   volume={475},
   ISSN={1365-2966},
   url={http://dx.doi.org/10.1093/mnras/stx3112},
   DOI={10.1093/mnras/stx3112},
   number={1},
   journal={Monthly Notices of the Royal Astronomical Society},
   publisher={Oxford University Press (OUP)},
   author={Pillepich, Annalisa and Nelson, Dylan and Hernquist, Lars and Springel, Volker and Pakmor, Rüdiger and Torrey, Paul and Weinberger, Rainer and Genel, Shy and Naiman, Jill P and Marinacci, Federico and Vogelsberger, Mark},
   year={2017},
   month=dec, pages={648–675} }

@ARTICLE{Sergio_2023,
       author = {{Contreras}, Sergio and {Chaves-Montero}, Jon{\'a}s and {Angulo}, Raul E.},
        title = "{Consistent clustering and lensing of SDSS-III BOSS galaxies with an extended abundance matching formalism}",
      journal = {\mnras},
         year = 2023,
        month = oct,
       volume = {525},
       number = {2},
        pages = {3149-3161},
          doi = {10.1093/mnras/stad2434},
archivePrefix = {arXiv},
       eprint = {2305.09637},
 primaryClass = {astro-ph.CO},
       adsurl = {https://ui.adsabs.harvard.edu/abs/2023MNRAS.525.3149C}
}

@ARTICLE{Sergio_2021,
       author = {{Contreras}, S. and {Angulo}, R.~E. and {Zennaro}, M.},
        title = "{A flexible modelling of galaxy assembly bias}",
      journal = {\mnras},
         year = 2021,
        month = jul,
       volume = {504},
       number = {4},
        pages = {5205-5220},
          doi = {10.1093/mnras/stab1170},
archivePrefix = {arXiv},
       eprint = {2005.03672},
 primaryClass = {astro-ph.GA},
       adsurl = {https://ui.adsabs.harvard.edu/abs/2021MNRAS.504.5205C}
}

@ARTICLE{Lacerna2025,
      title={Assessing the connection between galactic conformity and assembly-type bias}, 
      author={Ivan Lacerna and Nelson Padilla and Daniela Palma},
      year={2025},
      journal={arXiv:2505.03880},
      doi={10.48550/arXiv.2505.03880}
}

@article{Salcedo_2018,
   title={Spatial clustering of dark matter haloes: secondary bias, neighbour bias, and the influence of massive neighbours on halo properties},
   volume={475},
   ISSN={1365-2966},
   url={http://dx.doi.org/10.1093/mnras/sty109},
   DOI={10.1093/mnras/sty109},
   number={4},
   journal={Monthly Notices of the Royal Astronomical Society},
   publisher={Oxford University Press (OUP)},
   author={Salcedo, Andrés N and Maller, Ariyeh H and Berlind, Andreas A and Sinha, Manodeep and McBride, Cameron K and Behroozi, Peter S and Wechsler, Risa H and Weinberg, David H},
   year={2018},
   month=jan, pages={4411–4423} }

@article{Han_2018,
   title={The multidimensional dependence of halo bias in the eye of a machine: a tale of halo structure, assembly, and environment},
   volume={482},
   ISSN={1365-2966},
   url={http://dx.doi.org/10.1093/mnras/sty2822},
   DOI={10.1093/mnras/sty2822},
   number={2},
   journal={Monthly Notices of the Royal Astronomical Society},
   publisher={Oxford University Press (OUP)},
   author={Han, Jiaxin and Li, Yin and Jing, Yipeng and Nishimichi, Takahiro and Wang, Wenting and Jiang, Chunyan},
   year={2018},
   month=oct, pages={1900–1919} }

@article{Sato_Polito_2019,
   title={The dependence of halo bias on age, concentration, and spin},
   volume={487},
   ISSN={1365-2966},
   url={http://dx.doi.org/10.1093/mnras/stz1338},
   DOI={10.1093/mnras/stz1338},
   number={2},
   journal={Monthly Notices of the Royal Astronomical Society},
   publisher={Oxford University Press (OUP)},
   author={Sato-Polito, Gabriela and Montero-Dorta, Antonio D and Abramo, L Raul and Prada, Francisco and Klypin, Anatoly},
   year={2019},
   month=may, pages={1570–1579} }

@ARTICLE{MD2020,
       author = {{Montero-Dorta}, Antonio D. and {Artale}, M. Celeste and {Abramo}, L. Raul and {Tucci}, Beatriz and {Padilla}, Nelson and {Sato-Polito}, Gabriela and {Lacerna}, Ivan and {Rodriguez}, Facundo and {Angulo}, Raul E.},
        title = "{The manifestation of secondary bias on the galaxy population from IllustrisTNG300}",
      journal = {\mnras},
         year = 2020,
        month = aug,
       volume = {496},
       number = {2},
        pages = {1182-1196},
          doi = {10.1093/mnras/staa1624},
archivePrefix = {arXiv},
       eprint = {2001.01739},
 primaryClass = {astro-ph.GA},
       adsurl = {https://ui.adsabs.harvard.edu/abs/2020MNRAS.496.1182M}
}

@article{Xu_2020,
   title={Galaxy assembly bias of central galaxies in the Illustris simulation},
   volume={492},
   ISSN={1365-2966},
   url={http://dx.doi.org/10.1093/mnras/staa009},
   DOI={10.1093/mnras/staa009},
   number={2},
   journal={Monthly Notices of the Royal Astronomical Society},
   publisher={Oxford University Press (OUP)},
   author={Xu, Xiaoju and Zheng, Zheng},
   year={2020},
   month=jan, pages={2739–2754} }

@article{MD2021,
   title={On the influence of halo mass accretion history on galaxy properties and assembly bias},
   volume={508},
   ISSN={1365-2966},
   url={http://dx.doi.org/10.1093/mnras/stab2556},
   DOI={10.1093/mnras/stab2556},
   number={1},
   journal={Monthly Notices of the Royal Astronomical Society},
   publisher={Oxford University Press (OUP)},
   author={Montero-Dorta, Antonio D and Chaves-Montero, Jonás and Artale, M Celeste and Favole, Ginevra},
   year={2021},
   month=sep, pages={940–949} }

@article{Obuljen_2019,
   title={Anisotropic halo assembly bias and redshift-space distortions},
   volume={2019},
   ISSN={1475-7516},
   url={http://dx.doi.org/10.1088/1475-7516/2019/10/020},
   DOI={10.1088/1475-7516/2019/10/020},
   number={10},
   journal={Journal of Cosmology and Astroparticle Physics},
   publisher={IOP Publishing},
   author={Obuljen, Andrej and Dalal, Neal and Percival, Will J.},
   year={2019},
   month=oct, pages={020–020} }

@article{Yuan_2021,
   title={Evidence for galaxy assembly bias in BOSS CMASS redshift-space galaxy correlation function},
   volume={502},
   ISSN={1365-2966},
   url={http://dx.doi.org/10.1093/mnras/stab235},
   DOI={10.1093/mnras/stab235},
   number={3},
   journal={Monthly Notices of the Royal Astronomical Society},
   publisher={Oxford University Press (OUP)},
   author={Yuan, Sihan and Hadzhiyska, Boryana and Bose, Sownak and Eisenstein, Daniel J and Guo, Hong},
   year={2021},
   month=jan, pages={3582–3598} }

@ARTICLE{Pearl_2024,
       author = {{Pearl}, Alan N. and {Zentner}, Andrew R. and {Newman}, Jeffrey A. and {Bezanson}, Rachel and {Wang}, Kuan and {Moustakas}, John and {Aguilar}, Jessica N. and {Ahlen}, Steven and {Brooks}, David and {Claybaugh}, Todd and {Cole}, Shaun and {Dawson}, Kyle and {de la Macorra}, Axel and {Doel}, Peter and {Forero-Romero}, Jamie E. and {Gontcho A Gontcho}, Satya and {Honscheid}, Klaus and {Landriau}, Martin and {Manera}, Marc and {Martini}, Paul and {Meisner}, Aaron and {Miquel}, Ramon and {Nie}, Jundan and {Percival}, Will and {Prada}, Francisco and {Rezaie}, Mehdi and {Rossi}, Graziano and {Sanchez}, Eusebio and {Schubnell}, Michael and {Tarl{\'e}}, Gregory and {Weaver}, Benjamin A. and {Zhou}, Zhimin},
        title = "{The DESI One-percent Survey: Evidence for Assembly Bias from Low-redshift Counts-in-cylinders Measurements}",
      journal = {\apj},
         year = 2024,
        month = mar,
       volume = {963},
       number = {2},
          eid = {116},
        pages = {116},
          doi = {10.3847/1538-4357/ad1ffd},
archivePrefix = {arXiv},
       eprint = {2309.08675},
 primaryClass = {astro-ph.CO},
       adsurl = {https://ui.adsabs.harvard.edu/abs/2024ApJ...963..116P}
}

\appendix

% \section{Some extra material}

% If you want to present additional material which would interrupt the flow of the main paper,
% it can be placed in an Appendix which appears after the list of references.

%%%%%%%%%%%%%%%%%%%%%%%%%%%%%%%%%%%%%%%%%%%%%%%%%%

% Don't change these lines
\bsp	% typesetting comment
\label{lastpage}
\end{document}